\documentclass[aps,twocolumn,superscriptaddress,showpacs,showkeys]{revtex4}
\usepackage{graphics,graphicx,dcolumn,bm,fleqn,epic,eepic,float}
\usepackage{amssymb,amsmath,multirow,rotate,color,float,times}
\usepackage{color}
\usepackage{soul}                    
\definecolor{red}{rgb}{1,0,0}
\definecolor{green}{rgb}{0,1,0}
\definecolor{blue}{rgb}{0,0,1}

\newcommand{\bite}{\begin{itemize}}
\newcommand{\eite}{\end{itemize}}
\newcommand{\benu}{\begin{enumerate}}
\newcommand{\eenu}{\end{enumerate}}
\newcommand{\bverb}{\begin{verbatim}}
\newcommand{\everb}{\end{verbatim}}

\newcommand{\beq}{\begin{equation}}
\newcommand{\eeq}{\end{equation}}
\newcommand{\barr}{\begin{array}}
\newcommand{\earr}{\end{array}}

\begin{document}

\title{Extracting strong measurement noise from stochastic series:\\
       applications to empirical data}

\author{P.G.~Lind}
\affiliation{Center for Theoretical and Computational Physics, 
             University of Lisbon,
             Av.~Prof.~Gama Pinto 2, 1649-003 Lisbon, Portugal}
\affiliation{Departamento de F\'{\i}sica, Faculdade de Ci\^encias 
             da Universidade de Lisboa, 1649-003 Lisboa, Portugal} 
\author{M.~Haase}
\affiliation{Institute for High Performance Computing, University of Stuttgart,
             Nobelstr.~19, D-70569 Stuttgart, Germany}
\author{F.~B\"ottcher}
\affiliation{Institute of Physics, University of Oldenburg,
             D-26111 Oldenburg, Germany}
\author{J.~Peinke}
\affiliation{Institute of Physics, University of Oldenburg,
             D-26111 Oldenburg, Germany}
\author{D.~Kleinhans}
\affiliation{Institute for Marine Ecology, University of Gothenburg,
             Box 461, SE-405 30 G\"oteborg, Sweden}
\affiliation{Institute of Theoretical Physics, University of
             M\"unster, D-48149 M\"unster, Germany}
\author{R.~Friedrich}
\affiliation{Institute of Theoretical Physics, University of
             M\"unster, D-48149 M\"unster, Germany}

\begin{abstract}
It is a big challenge in the analysis of experimental data to disentangle 
the unavoidable measurement noise from the intrinsic dynamical noise.
Here we present a general operational method to extract measurement noise 
from stochastic time series, even in the case when the amplitudes of 
measurement noise and uncontaminated signal are of the same order 
of magnitude.
Our approach is based on a recently developed method for a nonparametric 
reconstruction of Langevin processes.
Minimizing a proper non-negative function the procedure is able to correctly 
extract strong measurement noise and to estimate
drift and diffusion coefficients in the Langevin equation describing the 
evolution of the original uncorrupted signal.
As input, the algorithm uses only the two first conditional moments 
extracted directly from the stochastic series and is therefore
suitable for a broad panoply of different signals.
To demonstrate the power of the method we apply the algorithm to synthetic
as well as climatological measurement data, namely the daily North Atlantic 
Oscillation index, shedding new light on the discussion of the nature of its 
underlying physical processes.
\end{abstract}

\pacs{05.40.Ca,  
      02.50.Ey,  
      92.70.Gt}  

\keywords{Measurement noise, Stochastic processes, Climate change}

\maketitle


\section{Introduction}
\label{sec:int}

Recently, much effort has been made to uncover
the dynamical process underlying a given time series of scale and time
dependent complex systems\cite{schreiberbook,friedrich08,abarbanel93}.
In many cases it is possible to describe such systems 
by a Langevin equation, extracted directly from the data, which separates
the deterministic and stochastic processes inherent to the 
system\cite{friedrich97}.
Such an approach has already been carried out successfully for instance
for data from turbulent fluid dynamics\cite{nawroth07}, financial 
data\cite{friedrich00}, climate indices\cite{collette04,lind05}
and for electroencephalographic recordings from epilepsy 
patients\cite{prusseit08,lehnertz09} and additional improvements were proposed 
to address the case of low sampling 
rates\cite{kleinhans05,gottschall08}.

However, typically the signal is subject to noise, due to 
experimental constraints or due to the measurement or discretization 
procedure leading to the data set to be studied.
Such noise is not intrinsic to the system, differing from what 
is known as dynamical noise, and therefore one is interested to 
separate it from the stochastic process.
We call such non-intrinsic noise measurement noise.
To separate the measurement noise from the dynamics of the measured
variable different predictor models or schemes for noise reduction
may be used\cite{schreiberbook,abarbanel93}.
In this context, an alternative procedure has been proposed\cite{boettcher06}
to extract the intrinsic dynamics associated with Langevin processes 
strongly contaminated by measurement noise, based solely on the two
conditional moments directly calculated
from the data\cite{gottschall08,boettcher06}.

In this manuscript we will revisit this nonparametric procedure, 
describing it in detail and explaining the main steps for its 
implementation, with the aim of applying it to empirical data sets.
Let us consider a one-dimensional Langevin process $x(t)$
(an extension to more dimensions is straightforward) defined as
\begin{equation}
\frac{dx}{dt} = D_1(x) + \sqrt{D_2(x)}\Gamma_t ,
\label{xlangevin}
\end{equation}
where $\Gamma_t$ represents a Gaussian $\delta$-correlated 
white noise 
$\langle\Gamma_t\rangle = 0$ and
$\langle\Gamma_t\Gamma_{t^{\prime}}\rangle = \delta(t-t^{\prime})$.
Functions $D_1(x)$ and $D_2(x)$ are the drift and diffusion 
coefficients defined as 
\begin{eqnarray}
D_n(x) = \frac{1}{n!}\lim_{\tau \to 0}\frac{1}{\tau}{M}_n(x,\tau)
\label{KramersMoyal}
\end{eqnarray}
for $n=1,2$, where ${M}_n(x,\tau)$ denotes the $n$-th order conditional 
moment of the data, as explained below.
Further, we consider that $x(t)$ is `contaminated' by a Gaussian
$\delta$-correlated measurement white noise, which leads to the series 
of observations
\begin{equation}
y(t) = x(t) + \sigma \zeta(t)
\label{yy}
\end{equation}
where $\sigma$ denotes the amplitude of the measurement noise.

When there is no measurement noise ($\sigma=0$), Eq.~(\ref{yy}) yields
the particular case $y(t)\equiv x(t)$, and the evolution
equation underlying the signal can be extracted directly from 
the two conditional moments ($n=1,2$)
\begin{eqnarray}
\hat{M}_n(y_i,\tau) &=& \langle (y(t+\tau)-y(t))^n \rangle |_{y(t)=y_i}
\label{gencondmom}  
\end{eqnarray}
as described in Refs.~\cite{friedrich97,friedrich00,lind05,boettcher06}.

In the presence of measurement noise ($\sigma\ne 0$) the conditional 
moments depend on $x$, $\tau$ and $\sigma$. Since generally
the limit
\begin{equation}
\lim\limits_{\tau\to 0} \hat{M}_n(x,\sigma\ne 0,\tau)
\end{equation}
does not exist, Eq.~(3) cannot be applied. 
The aim of this paper, however, is to explicitly derive a procedure 
which can transform the functional form of the 'noisy conditional 
moments' $\hat{M}_1(x,\sigma,\tau)$ and 
$\hat{M}_2(x,\sigma,\tau)$ at small $\tau$ 
into the 'true' coefficients $D_1(x)$ and $D_2(x)$ and simultaneously 
retrieve the amplitude $\sigma$ of the associated measurement noise.
For that, we show that $\hat{M}_n(y,\tau)$ for fixed $y$ is typically
linear in $\tau$ for a certain range $[\tau_1,\tau_2]$ of values 
(see Fig.~\ref{fig4} below). Therefore, even when $\sigma\neq 0$ one 
can estimate the quantities 
\begin{equation}
\hat{D_n}(y) = \frac{\hat{M}_n(y,\tau_2)- \hat{M}_n(y,\tau_1)}
                    {n!(\tau_2-\tau_1)} .
\label{Dprime}
\end{equation}

We start in Sec.~\ref{sec:processes} by briefly describing 
the procedure to extract Langevin equations from data sets 
and show how the drift and diffusion coefficients depend on the 
measurement noise strength $\sigma$. 
In particular, we will see that the proposed estimate\cite{siefert03} 
does not
yield the correct value when the measurement noise is too strong.
In Sec.~\ref{sec:algor} we then proceed to minimize a proper least
square function using the Levenberg-Marquardt procedure\cite{numrecip}.
By applying this algorithm to synthetic data we show that indeed
this approach is able to reliably extract the noise amplitude even in 
cases where it is of the same order as the synthetic signal without 
noise.
Furthermore, the procedure yields simultaneously more accurate estimates
for the clean signal $x(t)$.
Finally, in Sec.~\ref{sec:applications}, we apply this framework to 
an empirical data set, namely the North Atlantic Oscillation daily
index\cite{ijbc}, giving some insight from the obtained results to 
the underlying system.
Discussion and conclusions are given in Sec.~\ref{sec:conc}, where further
possible applications are proposed.
All details concerning the implementation of the minimization
procedure to extract strong measurement noise are given as appendices.

\section{Stochastic time series with strong measurement noise}
\label{sec:processes}

We consider a time series generated by integrating Eq.~(\ref{xlangevin}) 
with drift and diffusion coefficient assumed to be linear and quadratic
forms respectively
\begin{subequations}
\begin{eqnarray}
D_1(x) &=& d_{10} + d_{11}x \label{D1}\\
       & & \cr
D_2(x) &=& d_{20} + d_{21}x + d_{22}x^2 , \label{D2}
\end{eqnarray}\label{DD}\end{subequations}
and by adding separately to each data point the measurement 
term $\sigma\zeta(t)$ in Eq.~(\ref{yy}). 
Though we concentrate on the particular expressions for $D_1$ and $D_2$
given above, it should be stressed that they comprehend a large collection 
of different processes, such as Ornstein-Uhlenbeck processes\cite{boettcher06}.
Further, some generalizations may be carried out as will be discussed in
Sec.~\ref{sec:conc}.
Using Eqs.~(\ref{DD}a) and (\ref{DD}b), one has six parameters: five 
coefficients $d_{ij}$ defining the evolution equation of the clean signal 
and a sixth parameter $\sigma$ for the amplitude of the measurement noise.
\begin{figure}[htb]
\begin{center}
\includegraphics*[width=8.0cm,angle=0]{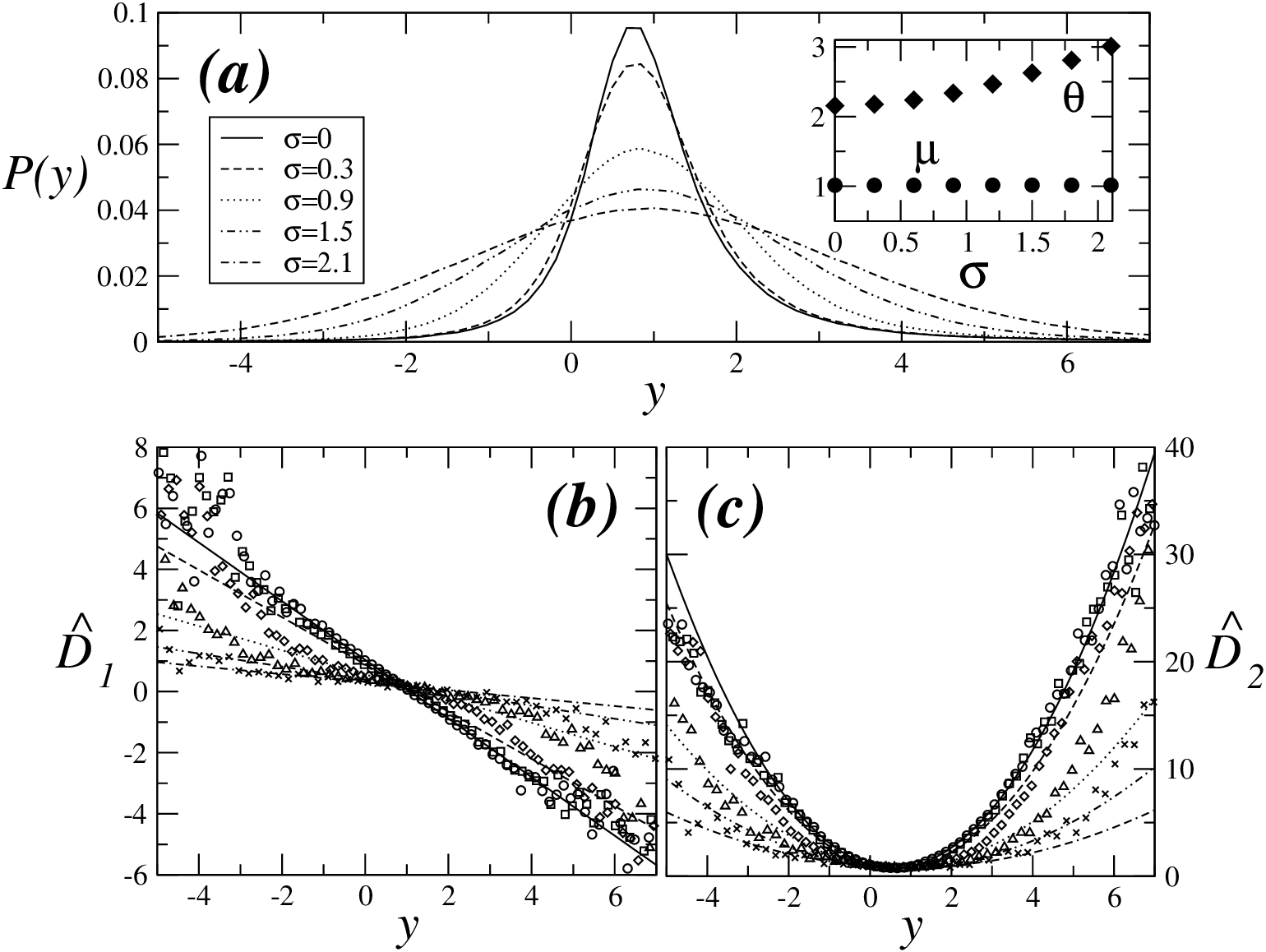}
\end{center}
\caption{\protect Langevin time series with different measurement
  noise strengths. Here we show 
  {\bf (a)} the probability density function $P(y)$ of the series
  with noise (see Eq.~(\ref{yy})), with the corresponding mean value 
  $\mu$ and standard deviation $\theta$ in the inset, and the 
  corresponding functions
  {\bf (b)} ${\hat D}_1(y)$ and
  {\bf (c)} ${\hat D}_2(y)$, see Eq.~(\ref{Dprime}).
  In all cases, the assumed time series $x(t)$ without measurement
  noise uses the coefficients $D_1(x)=1-x$ and $D_2(x)=1-x+x^2$.}
\label{fig1}
\end{figure}
\begin{figure}[htb]
\begin{center}
\includegraphics*[width=8.0cm,angle=0]{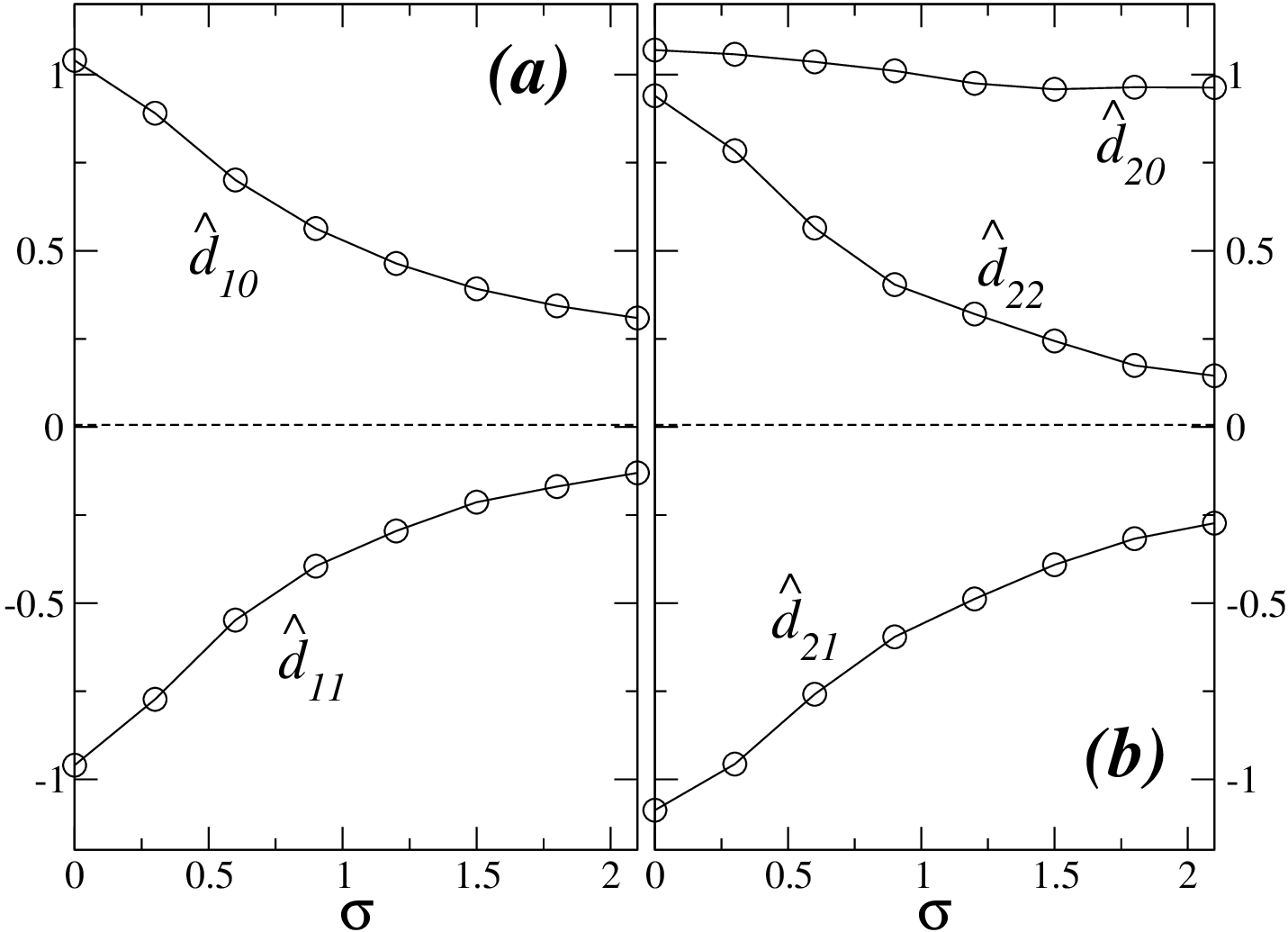}
\end{center}
\caption{\protect 
  Noise dependence of functions $\hat{D}_1(y)$ and $\hat{D}_2(y)$ 
  (see text and Eq.~(\ref{Dprime}))
  The underlying Langevin time series $x(t)$ without noise 
  is the same as in Fig. \ref{fig1}.}
\label{fig2}
\end{figure}

Figure \ref{fig1} illustrates this influence of noise for a particular 
choice of $D_1(x),D_2(x)$.
As shown in Fig.~\ref{fig1}a, for increasing $\sigma$ one obtains
broader probability density functions $P(y)$ as one intuitively
expects.
Quantitatively, the standard deviation $\theta$ of $P(y)$ varies 
quadratically with the measurement noise $\sigma$, while the mean value
$\mu$ of $P(y)$ remains constant, as shown in the inset of Fig.~\ref{fig1}a.
The estimated functions $\hat{D}_1(y)$ and $\hat{D}_2(y)$ 
change significantly, as shown in Fig.~\ref{fig1}b and \ref{fig1}c 
respectively.
Assuming ${\hat D}_1(y)={\hat d}_{10}+{\hat d}_{11}y$ and 
${\hat D}_2(y)={\hat d}_{20}+{\hat d}_{21}y+{\hat d}_{22}y^2$, 
Fig.~\ref{fig2} shows how the estimated parameters $\hat{d}_{ij}$ deviate 
from the `true' uncontaminated values $d_{ij}$ in Eq.~(\ref{DD})
when measurement noise increases.
Notice that for $\sigma=0$ -- see left vertical axis in the plots of 
Fig.~\ref{fig2} -- the estimated parameter values are approximately 
correct.
\begin{figure}[t]
\begin{center}
\includegraphics*[width=8.5cm,angle=0]{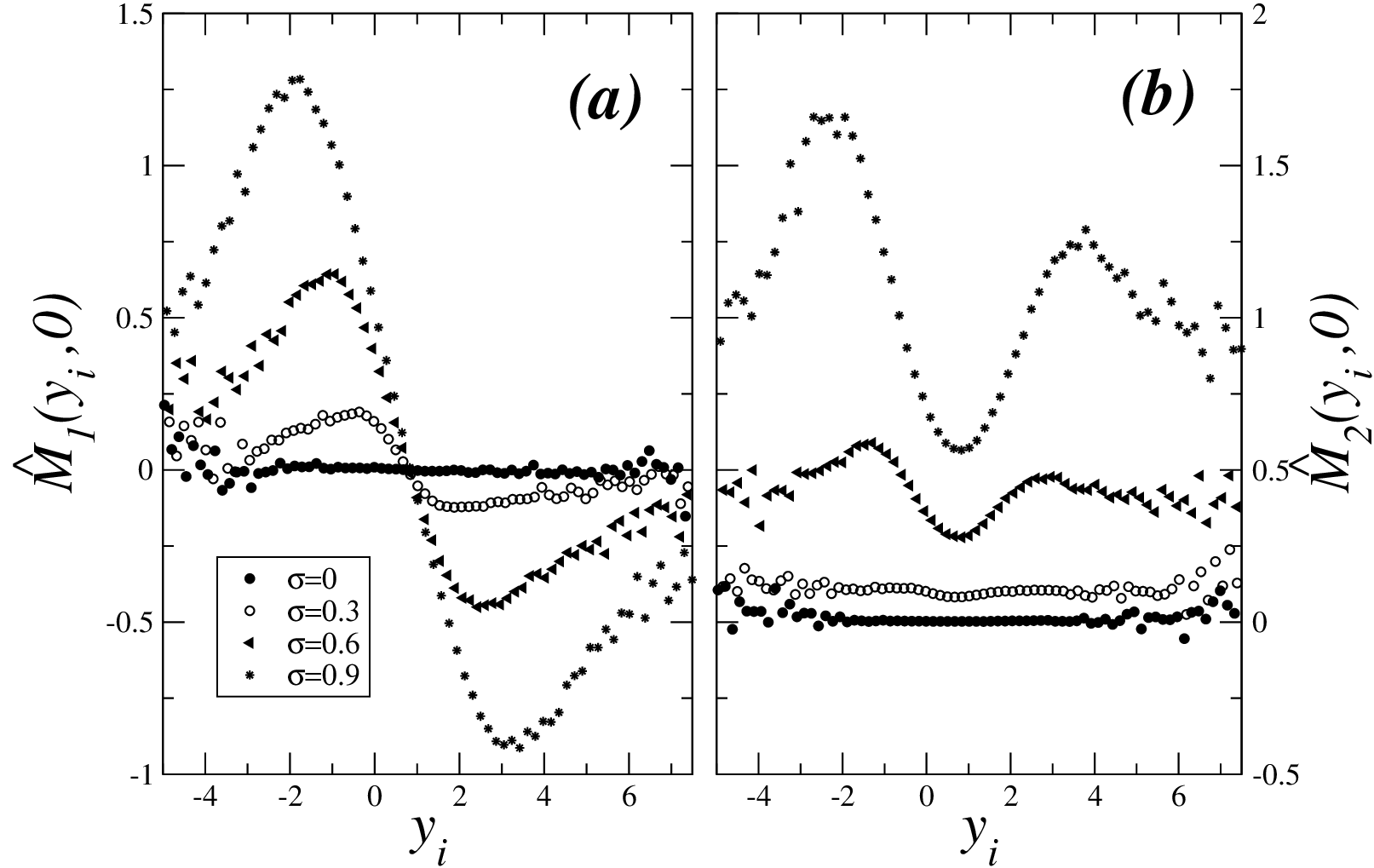}
\end{center}
\caption{\protect  
   Conditional moments ${\hat M}_1(y_i,\tau)$ and ${\hat M}_2(y_i,\tau)$ 
   as a function of bin $y_i$, for $\tau=0$ and different measurement 
   noise strengths. The asymmetry of ${\hat M}_2$ is due to $d_{21}\neq 0$
   (see Eqs.~(\ref{DD})).
   The same $x(t)$ as in Fig.~\ref{fig1} was used.}
\label{fig3}
\end{figure}
\begin{figure}[htb]
\begin{center}
\includegraphics*[width=8.5cm,angle=0]{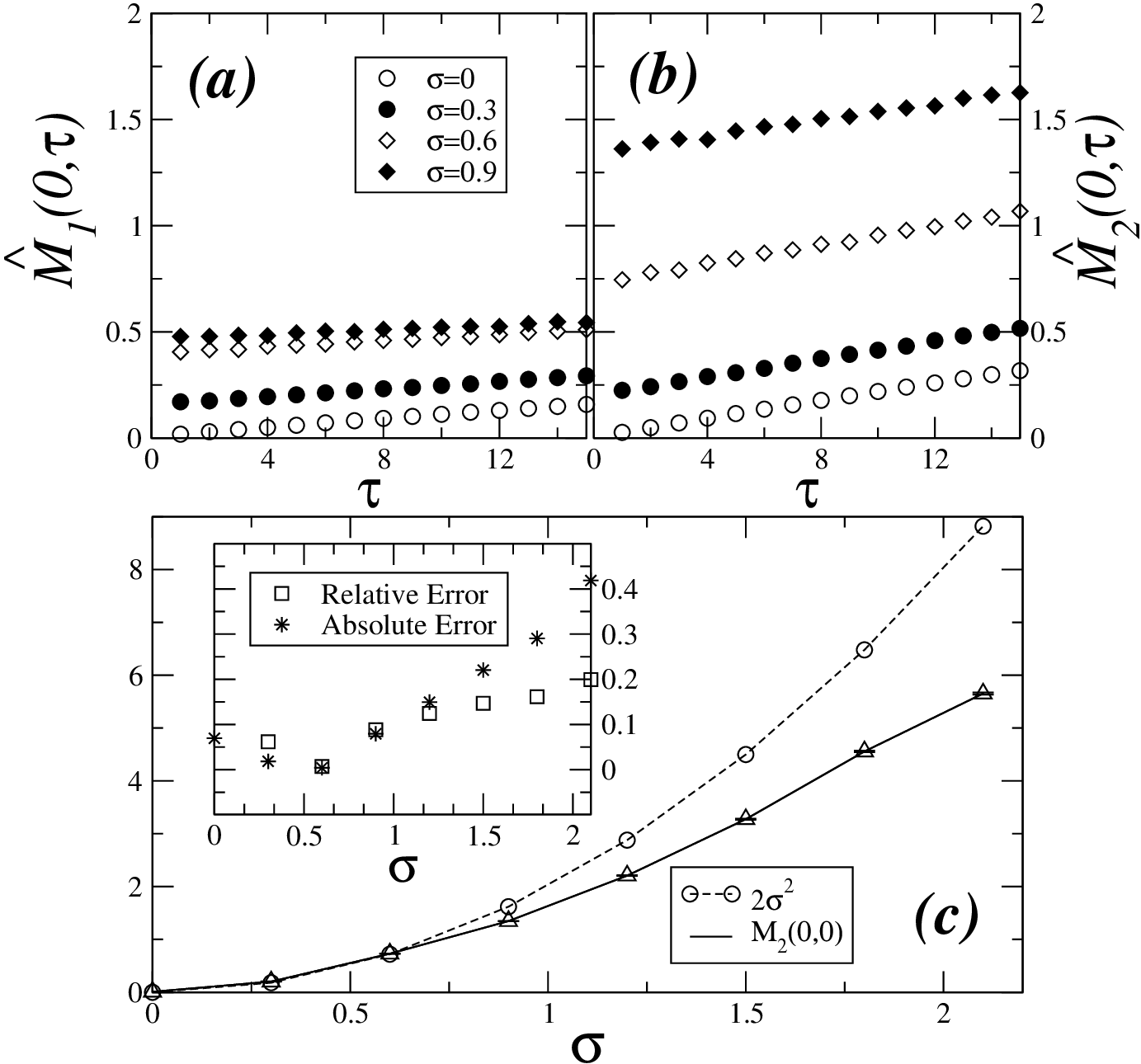}
\end{center}
\caption{\protect   
   Conditional moments 
   {\bf (a)} ${\hat M}_1(y_i,\tau)$ and 
   {\bf (b)} ${\hat M}_2(y_i,\tau)$ as a function of $\tau$, for
   bin $y_i=0$ and different measurement noise strengths. 
   In {\bf (c)} one compares the true measurement noise
   with the approximation $\sigma_{app}={\hat M}_2(0,0)\sim 2\sigma^2$ 
   given in Eq.~(\ref{approxmeasnoise}).
   In the inset the corresponding absolute and relative erros are given
   by $\zeta_a=\vert \sigma-\sigma_{app}\vert$ and $\zeta_r=\zeta_a/\sigma$
   respectively. Errors for $\hat{M}_2$ are negligible.
   The same $x(t)$ as in Fig.~\ref{fig1} was used.}
\label{fig4}
\end{figure}
\begin{figure}[t]
\begin{center}
\includegraphics*[width=8.5cm,angle=0]{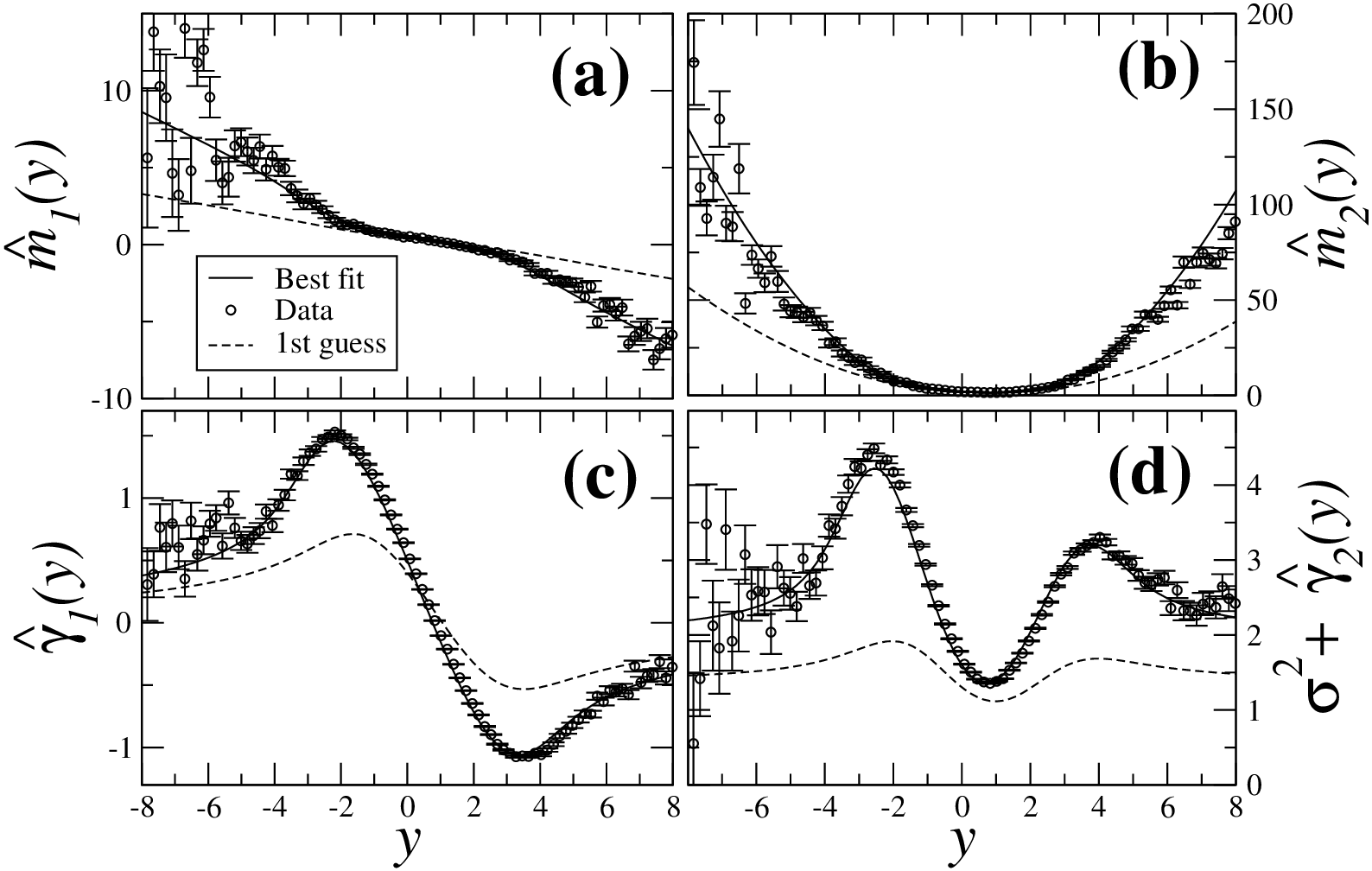}
\end{center}
\caption{\protect   
   Functions $\hat{m}_1$, $\hat{m}_2$, $\hat{\gamma}_1$ and
   $\hat{\gamma}_2$ (symbols) defining the conditional moments in 
   Eqs.~(\ref{yM_2}).
   The underlying Langevin time series $x(t)$ without noise is 
   characterized by a drift coefficient $D_1(x)=1-x$ and
   a diffusion coefficient $D_2(x)=1-x+x^2$.
   The measurement noise was fixed at $\sigma=1$.
   Each hat-function is compared with the corresponding
   integral form in Eqs.~(\ref{gamma-m}) using the first estimate
   of parameters values (dashed lines) and the true values (solid lines).}
\label{fig5}
\end{figure}

To correctly derive the drift and diffusion coefficients $D_1(x)$ and
$D_2(x)$ when $\sigma$ is strong, we consider the measured conditional
moments $\hat{M}_1(y_i,\tau)$ and $\hat{M}_2(y_i,\tau)$, as in 
Eq.~(\ref{gencondmom}), the hat indicating that they are calculated 
from the measured data $y(t)$ directly.
Since this conditional moments depend in a non trivial way on both 
time $\tau$ and amplitude $y_i$, we approximate them up to first 
order on $\tau$:
\begin{subequations}
\begin{eqnarray}
\hat{M}_1(y_i,\tau) &=& \langle y(t+\tau)-y(t) \rangle |_{y(t)=y_i}\cr
   & & \cr
   &=& \tau \hat{m}_1(y_i) + \hat{\gamma}_1(y_i) + {\cal O}(\tau^2), 
            \label{yM1_2}\\
   & & \cr
\hat{M}_2(y_i,\tau) &=& \langle (y(t+\tau)-y(t))^2 \rangle |_{y(t)=y_i}\cr
   & & \cr
   &=& \tau \hat{m}_2(y_i) + \hat{\gamma}_2(y_i) + \sigma^2 + 
            {\cal O}(\tau^2),
\label{yM2_2}
\end{eqnarray}
\label{yM_2}
\end{subequations}
where $y(t)$ is taken in the range $y_i\pm\Delta y/2$ for
each bin $i$, and $\Delta y$ depends on the binning considered.
Appendix \ref{app:km} gives the full derivation of Eqs.~(\ref{yM_2}).

Figure \ref{fig3} shows both conditional moments for $\tau=0$ and
with different measurement noise strengths. 
Conversely, in Fig.~\ref{fig4}a and \ref{fig4}b one sees that the 
conditional moments depend linearly on $\tau$ for a fixed amplitude 
$y$, which justifies the approximation assumed in Eqs.~(\ref{yM_2}).
Therefore, to study the dependence of the conditional moments on $y$ 
we will consider the linear decompositions in Eqs.~(\ref{yM_2}), as 
done in Fig.~\ref{fig5}.
Our simulations with synthetic data have shown that using a to
large range of $\tau$ values yields results for $D_1$ and $D_2$
deviated from their true values. The best estimation for both
Kramers-Moyal coefficients are obtained using the range $1 < \tau 
\lesssim 4$.

Notice that for sufficiently
small measurement noise a good estimate of it is given 
by\cite{boettcher06,siefert03}
\begin{equation}
\sigma \approx \sqrt{\frac{\hat{M}_2(\mu,0)}{2}},
\label{approxmeasnoise}
\end{equation}
where $\mu$ is the average value of $y(t)$ data points in the 
time series. For details see Append.~\ref{app:km}.
However, as shown in Fig.~\ref{fig4}c, this approximation is no longer
valid for sufficiently high measurement noise,
namely when $\sigma \gtrsim 0.5$ (see inset of Fig.~\ref{fig4}c)
and even otherwise coefficients $D_1$ and $D_2$ are not correctly estimated 
(see Fig.~\ref{fig2}). 
Therefore, a better algorithm to estimate such parameters is necessary.

The heart of our procedure to correctly estimate measurement noise
lies in the fact that while the functions $\hat{m}_i$ and 
$\hat{\gamma}_i$ (i=1,2) are obtained explicitly for each
bin value $y_i$, functions $m_i$ and $\gamma_i$ depend 
generally on the drift and diffusion coefficients as follows:
\begin{subequations}
\begin{eqnarray}
\gamma_1(y) &=& \int_{-\infty}^{+\infty} 
                (x-y)\bar{f}_{\sigma}(x\vert y)dx \label{gamma1}\\
            & & \cr
\gamma_2(y) &=& \int_{-\infty}^{+\infty} 
                (x-y)^2 \bar{f}_{\sigma}(x\vert y)dx \label{gamma2}\\
            & & \cr
m_1(y)      &=& \int_{-\infty}^{+\infty} 
                 D_1(x) \bar{f}_{\sigma}(x\vert y)dx \label{m1}\\
            & & \cr
m_2(y)      &=& 2\int_{-\infty}^{+\infty} 
                [(x-y)D_1(x)+D_2(x)] \bar{f}_{\sigma}(x\vert y)dx , \cr
  & & \label{m2}
\end{eqnarray}\label{gamma-m}\end{subequations}
where $\bar{f}_{\sigma}(x\vert y)$ is the probability for the system to adopt the
value $x$ when a measured value $y$ is observed.
For details about the derivation of functions in Eqs.~(\ref{gamma-m})
see Append.~\ref{app:km} and for the explicit expression of 
$\bar{f}_{\sigma}(x\vert y)$ see Append.~\ref{app:condprob}.

In Fig.~\ref{fig5} we illustrate both the hat-functions in Eqs.~(\ref{yM_2})
and their integral form in Eqs.~(\ref{gamma-m}).
Due to the measurement noise fixed in this example at $\sigma=1$ the
hat-functions (symbols) are not properly fitted by the integral form
in Eqs.~(\ref{gamma-m}) 
using the first estimate (dashed lines) of the parameters $d_{ij}$,
taken from Fig.~\ref{fig2}, and $\sigma$, computed from 
Eq.~(\ref{approxmeasnoise}).
If instead we use the true parameter values in the integral forms of
our $m_i$ and $\gamma_i$ functions a proper fit is obtained (solid lines).
\begin{figure}[t]
\begin{center}
\includegraphics*[width=8.3cm,angle=0]{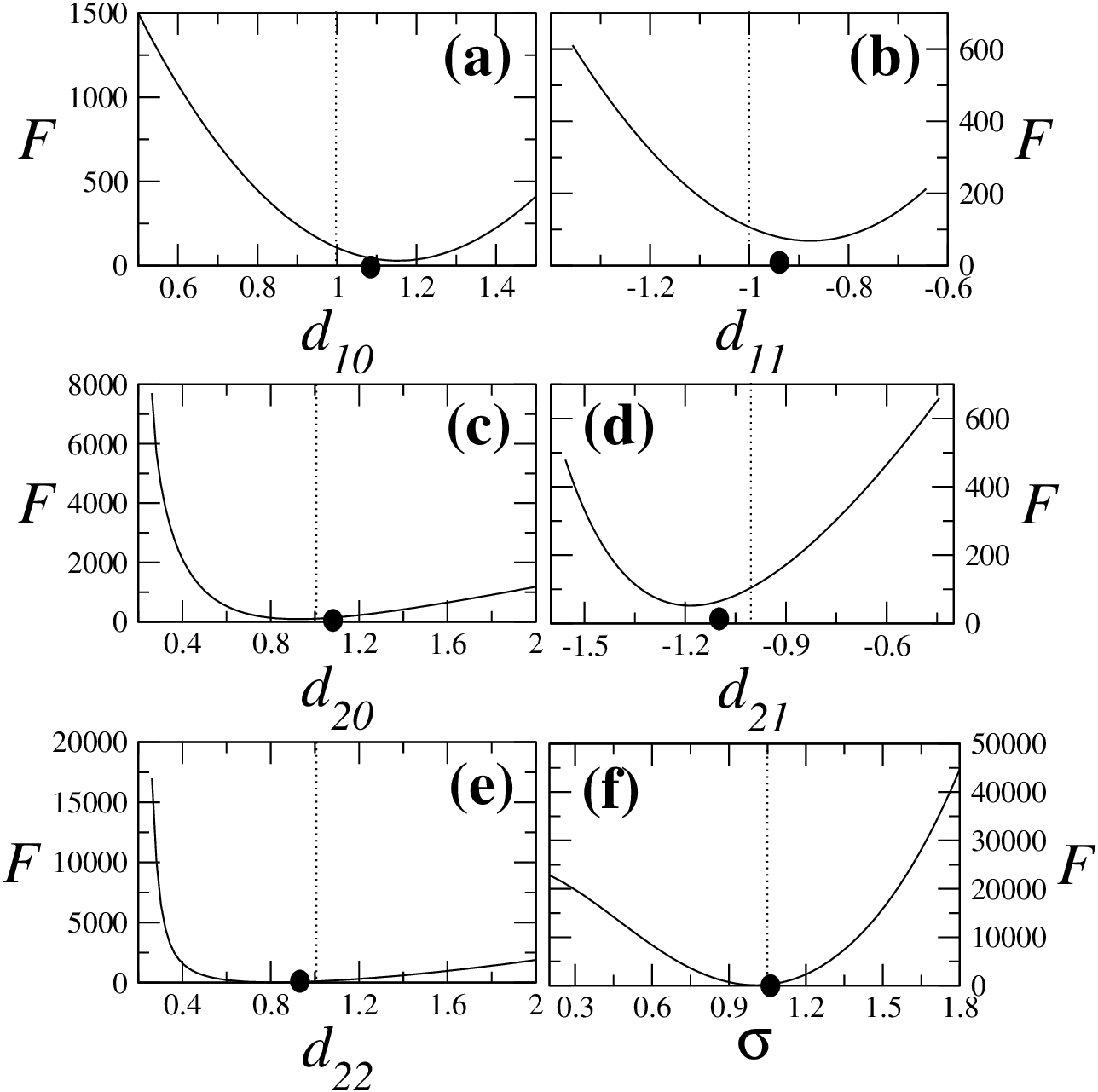}
\end{center}
\caption{\protect   
      Function $F$ in Eq.~(\ref{F}) as a function of 
      {\bf (a)} $d_{10}$,
      {\bf (b)} $d_{11}$,
      {\bf (c)} $d_{20}$,
      {\bf (d)} $d_{21}$,
      {\bf (e)} $d_{22}$ and
      {\bf (f)} $\sigma$.
      The same situation as in Fig.~\ref{fig2} is here chosen:
      $D_1(x)=1-x$, $D_2(x)=1-x+x^2$ and $\sigma=1$.
      Dashed lines indicate the true values used for generating the
      data series, while the bullet indicates the estimated values
      of the Kramers-Moyal coefficients for $\sigma=0$.
      In each plot while varying one parameter, the remaining ones
      are fixed at their true values (see text).}
\label{fig6}
\end{figure}

Therefore, the problem we want to solve is to determine the parameters
that minimize the function:
\begin{eqnarray}
F &=& \frac{1}{M}\sum_{i=1}^{M} \Big [ 
               \frac{\left ( \hat{\gamma}_1-\gamma_1(y_i)\right )^2}
                    {\sigma^2_{\hat{\gamma}_1}(y_i)} + \cr
   & & \cr
   & & \phantom{\frac{1}{M}\sum_{i=1}^{M} \Big [}
               \frac{\left ( \hat{\gamma}_2-\gamma_2(y_i)-\sigma^2\right )^2}
                    {\sigma^2_{\hat{\gamma}_2}(y_i)} + \cr
   & & \cr
   & & \phantom{\frac{1}{M}\sum_{i=1}^{M} \Big [}
               \frac{\left ( \hat{m}_1-m_1(y_i)\right )^2}
                    {\sigma^2_{\hat{m}_1}(y_i)} +\cr
   & & \cr
   & & \phantom{\frac{1}{M}\sum_{i=1}^{M} \Big [}
               \frac{\left ( \hat{m}_2-m_2(y_i)\right )^2}  
                    {\sigma^2_{\hat{m}_2}(y_i)} \Big ] ,
\label{F}
\end{eqnarray}
where the summation extends over all $M$ bins, $\sigma_{\hat{\gamma}_1}(y_i)$
is the error associated to function $\hat{\gamma}_1$ at the value
$y_i$ and similarly for $\sigma_{\hat{\gamma}_2}$, $\sigma_{\hat{m}_1}$
and $\sigma_{\hat{m}_2}$.
Notice that the values of such $\sigma_{\hat{\gamma}_i}$ and 
$\sigma_{\hat{m}_i}$ are taken directly from the data only.
See Appendix \ref{app:km} for details.

Taking again the example illustrated in Fig.~(\ref{fig2}) with
$\sigma=1$ we plot in Fig.~\ref{fig6} function $F$ in Eq.~(\ref{F})
as function of each one of the parameters keeping all others fixed
at their true values.
Evidently, the estimated values are near the minimum of $F$ in each
case. Further, the one-dimensional cuts of function $F$ show only one 
minimum. One should note however that, for the entire $6$-dimensional 
parameter space, several local minima of $F$ may appear. 
In fact, after minimizing $F$ by varying one parameter, function $F$ also 
changes as a function of the other parameters, i.e.~its minimum as a 
function of the other parameter changes. 
In the next Section we will see how to minimize function $F$, in order to 
find good estimates for the correct values for each parameter.

\section{Optimization procedure}
\label{sec:algor}

After computing the functions $\hat{\gamma}_1$, $\hat{\gamma}_2$, 
$\hat{m}_1$ and $\hat{m}_2$ as well as the
corresponding errors $\sigma_{\hat{\gamma}_1}$, etc, directly from 
the measured time series $y(t)$
and estimating the coefficients $D_1$ and $D_2$ given by the functional
forms in Eqs.~(\ref{DD})
there are several ways to minimize $F$.
All of them start from the initially estimated set of values for the 
parameters and iteratively improve the solution, by finding lower values 
of $F$, till convergence is attained. 

To proceed the following remark should be considered.
Parameter $d_{10}$ can be always eliminated with a simple
transformation $x\to x^{\prime}=x+d_{10}/d_{11}$. 
Alternatively, and since we do not know beforehand the
true values of $d_{10}$ and $d_{11}$ we can consider also the fact
that averaging Eq.~(\ref{xlangevin}) yields
$d_{10}=-d_{11}\langle x\rangle$ and consider the transformation
$x^{\prime}=x-\langle x\rangle$. 
With these arguments, we henceforth disregard $d_{10}$, which reduces the
dimension of parameter space by one. Parameter $d_{10}$ is computed
from the relations above, only after minimizing $F$.
For simplicity the primes in $x^{\prime}$ will be omitted.

The simplest way is to minimize each term in $F$ and repeat that a large 
number of times starting from different initial conditions for the 
parameters, in a sort of a Monte Carlo procedure of random 
walks~\cite{bouchaud90} or L\'evy-walks\cite{metzler00}.
The Monte Carlo procedure assures that a substantial number of local 
minimal for $F$ will be visited, and in the end we take the minimum 
of all $F$ values found.
Simulations have shown however that a Monte Carlo procedure is too 
expensive in this case, since there are different local minima and 
the choice of the minimum is strongly path dependent.
We will therefore consider the Levenberg-Marquardt method\cite{numrecip}. 

For the Levenberg-Marquardt procedure one computes the first and second 
derivative of $F$. Symbolizing the parameters $\sigma, d_{11}, d_{20}, d_{21}$ 
and $d_{22}$ by $p_k$ with $k=1,\dots,5$ respectively, these derivatives read
\begin{widetext}
\begin{eqnarray}
\frac{\partial F}{\partial p_k} &=& 
        -\frac{2}{M} \sum_{i=1}^M \Big [ 
               \frac{\hat{\gamma}_1-\gamma_1}
                    {\sigma^2_{\hat{\gamma}_1}(i)} 
               \frac{\partial \gamma_1}{\partial p_k} +
               \frac{\hat{\gamma}_2-\gamma_2-\sigma^2}
                    {\sigma^2_{\hat{\gamma}_2}(i)} 
               \frac{\partial (\gamma_2+\sigma^2)}{\partial p_k} +
               \frac{\hat{m}_1-m_1}
                    {\sigma^2_{\hat{m}_1}(i)} 
               \frac{\partial m_1}{\partial p_k} +
               \frac{\hat{m}_2-m_2}
                    {\sigma^2_{\hat{m}_2}(i)} 
               \frac{\partial m_2}{\partial p_k} \Big ] ,\label{derF}\\
   & & \cr
   & & \cr
\frac{\partial^2 F}{\partial p_k\partial p_{\ell}} &=& 
         \frac{2}{M} \sum_{i=1}^M \Big [ 
               \frac{1}{\sigma^2_{\hat{\gamma}_1}(i)}
               \frac{\partial \gamma_1}{\partial p_k} 
               \frac{\partial \gamma_1}{\partial p_{\ell}}- 
               \frac{\hat{\gamma}_1-\gamma_1}
                    {\sigma^2_{\hat{\gamma}_1}(i)} 
               \frac{\partial^2 \gamma_1}{\partial p_k\partial p_{\ell}} +\cr
  & & \cr
  & & \cr
  & & \phantom{\frac{2}{M} \sum_{i=1}^M \Big [}
               \frac{1}{\sigma^2_{\hat{\gamma}_2}(i)}
               \frac{\partial (\gamma_2+\sigma^2)}{\partial p_k} 
               \frac{\partial (\gamma_2+\sigma^2)}{\partial p_{\ell}}- 
               \frac{\hat{\gamma}_2-\gamma_2-\sigma^2}
                    {\sigma^2_{\hat{\gamma}_2}(i)} 
               \frac{\partial^2 (\gamma_2+\sigma^2)}
                    {\partial p_k\partial p_{\ell}} +\cr
  & & \cr
  & & \cr
  & & \phantom{\frac{2}{M} \sum_{i=1}^M \Big [}
               \frac{1}{\sigma^2_{\hat{m}_1}(i)}
               \frac{\partial m_1}{\partial p_k} 
               \frac{\partial m_1}{\partial p_{\ell}}- 
               \frac{\hat{m}_1-m_1}
                    {\sigma^2_{\hat{m}_1}(i)} 
               \frac{\partial^2 m_1}
                    {\partial p_k\partial p_{\ell}} +
               \frac{1}{\sigma^2_{\hat{m}_2}(i)}
               \frac{\partial m_2}{\partial p_k} 
               \frac{\partial m_2}{\partial p_{\ell}}- 
               \frac{\hat{m}_2-m_2}
                    {\sigma^2_{\hat{m}_2}(i)} 
               \frac{\partial^2 m_2}
                    {\partial p_k\partial p_{\ell}} \Big ] \cr
  & & \cr
  & & \cr
  &\sim& 
         \frac{2}{M} \sum_{i=1}^M \Big [ 
               \frac{1}{\sigma^2_{\hat{\gamma}_1}(i)}
               \frac{\partial \gamma_1}{\partial p_k} 
               \frac{\partial \gamma_1}{\partial p_{\ell}}
               +
               \frac{1}{\sigma^2_{\hat{\gamma}_2}(i)}
               \frac{\partial (\gamma_2+\sigma^2)}{\partial p_k} 
               \frac{\partial (\gamma_2+\sigma^2)}{\partial p_{\ell}}
               +\cr
  & & \cr
  & & \cr
  & & \phantom{\frac{2}{M} \sum_{i=1}^M \Big [}
               \frac{1}{\sigma^2_{\hat{m}_1}(i)}
               \frac{\partial m_1}{\partial p_k} 
               \frac{\partial m_1}{\partial p_{\ell}}
               +
               \frac{1}{\sigma^2_{\hat{m}_2}(i)}
               \frac{\partial m_2}{\partial p_k} 
               \frac{\partial m_2}{\partial p_{\ell}}
                -2\delta_{\sigma p_k}\delta_{\sigma p_{\ell}}
               \frac{\hat{\gamma}_2-\gamma_2-\sigma^2}
                    {\sigma^2_{\hat{\gamma}_2}(i)} 
                \Big ] .\label{derderF}
\end{eqnarray}
\end{widetext}
In the right-hand side of Eq.~(\ref{derderF}) we neglect the terms containing 
second derivatives of $\gamma$ and $m$ functions.
This last approximation of neglecting second derivatives is 
acceptable as far as the model is successful\cite{numrecip}. 

By symbolizing first and second derivatives as $\beta_k$ and
$\alpha_{k\ell}$ respectively the iterative procedure computes
the increments $dp_k$ for each parameter $p_k$ ($k=1,\dots,5$),
which are the solutions of
\begin{equation}
\beta_k=-\sum_{\ell=1}^5 \alpha_{k\ell}dp_{\ell} .
\label{betaalpha}
\end{equation}
Furthermore,
one assumes that $dp_{\ell}\propto \beta_{\ell}$, which considering dimensional
analysis\cite{numrecip} can be written as:
\begin{equation}
dp_{\ell}=\frac{\beta_{\ell}}{\lambda\alpha_{\ell\ell}} ,
\end{equation}
where typically $\lambda\gg 1$.
For a given $\lambda$ value, instead of the second derivatives $\alpha_{mn}$ 
one assumes $\alpha^{\prime}_{mn}=\alpha_{mn}(1+\lambda)$ for $m=n$ and 
$\alpha^{\prime}_{mn}=\alpha_{mn}$ otherwise and solves Eq.~(\ref{betaalpha}) 
for $dp_{k}$ \cite{numrecip}. 

If $F(p_k+dp_k) < F(p_k)$, the parameter values are updated, 
$p_k\to p_k+dp_k$, and $\lambda$ is typically decreased by $10\%$. 
Otherwise, if $F(p_k+dp_k) \ge F(p_k)$ one increases $\lambda$ by $10\%$ and 
determines new increments $dp_k$.
The procedure stops after attaining the required convergence.

\begin{figure}[htb]
\begin{center}
\includegraphics*[width=8.0cm,angle=0]{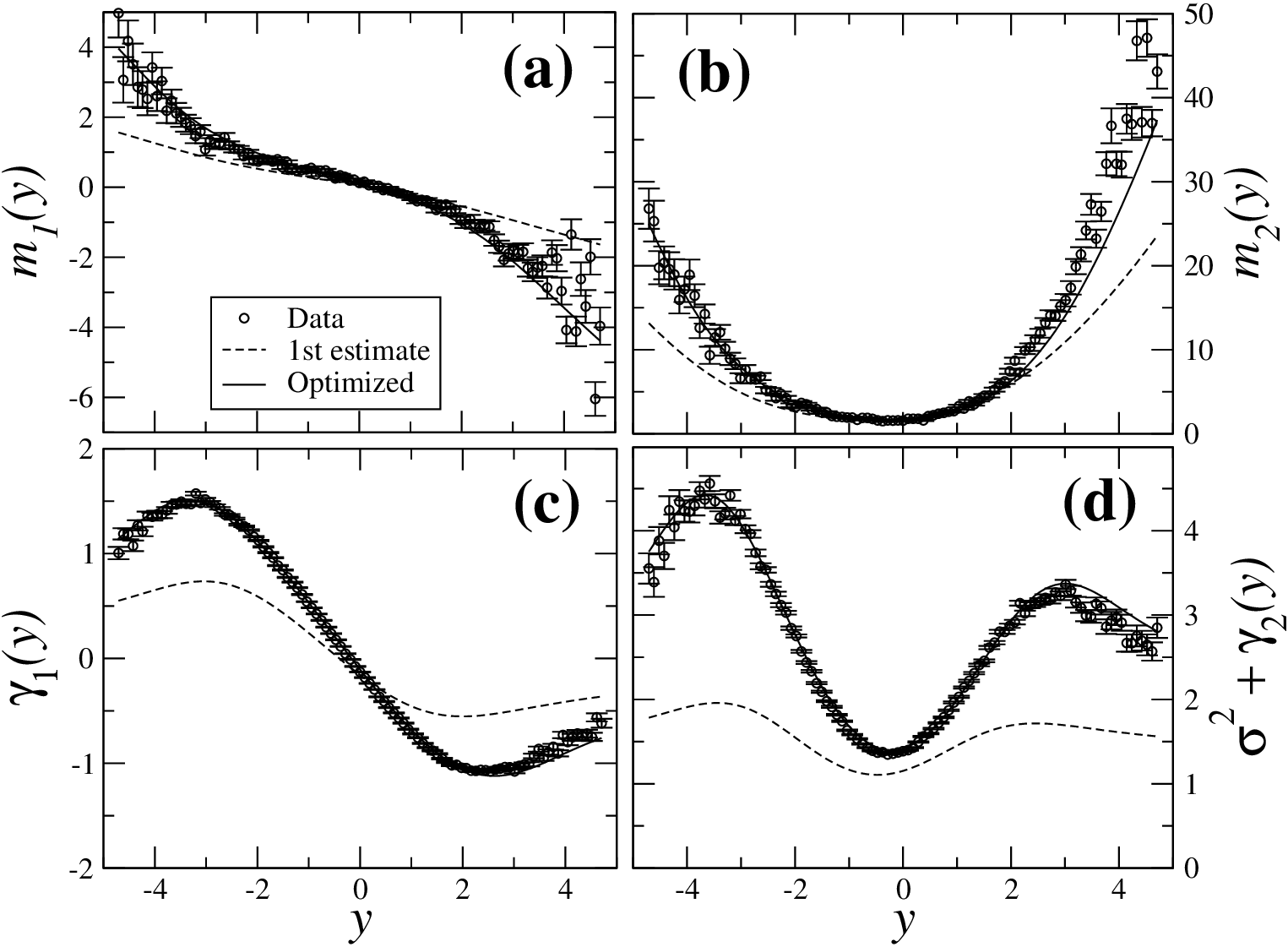}
\end{center}
\caption{\protect
        Functions $m_1$, $m_2$, $\gamma_1$ and $\gamma_2$
        for the Langevin process with $D_1(x)=1-x$,
        $D_2(x)=1-x+x^2$ and $\sigma=1$. Symbols indicate the
        functions obtained for the data, dashed line corresponds
        to the first estimate of the parameters and solid line
        corresponds to the parameter values obtained from the 
        Levenberg-Marquardt procedure (see text).
        In this case, for the first estimate one has $F_0=3720$
        while the final estimate retrieves $F_{LM}=33.1$.
        The true minimum is $F_m=29.2$.}
\label{fig7}
\end{figure}
\begin{figure}[htb]
\begin{center}
\includegraphics*[width=8.5cm,angle=0]{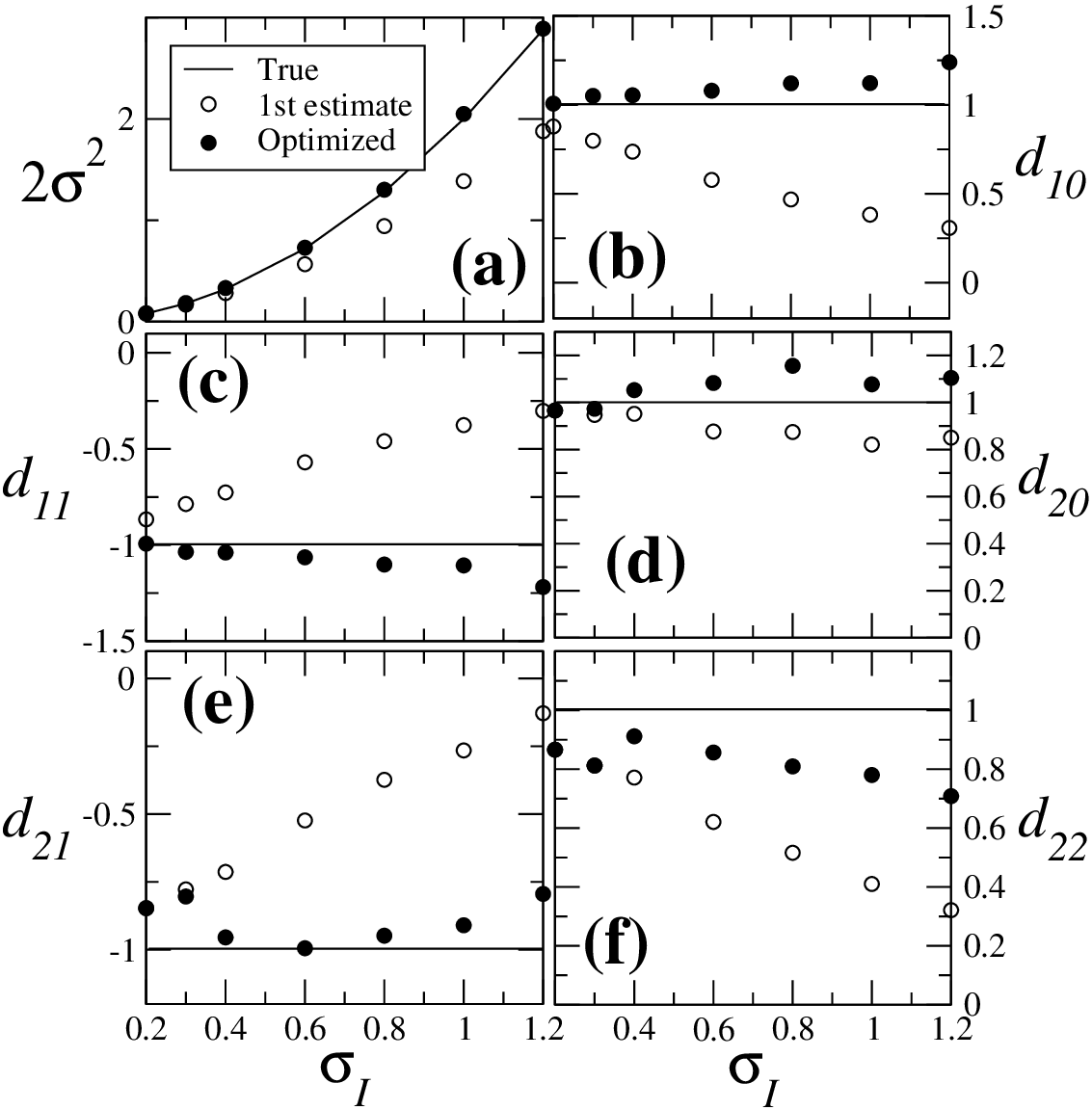}
\end{center}
\caption{\protect
        Comparison of the optimized parameters values (bullet)
        with the first estimate and the true values for different
        input measurement noise strengths $\sigma_I$:
        {\bf (a)} $2\sigma^2$,
        {\bf (b)} $d_{10}$,
        {\bf (c)} $d_{11}$,
        {\bf (d)} $d_{20}$,
        {\bf (e)} $d_{21}$,
        {\bf (f)} $d_{22}$.
        The measurement noise is correctly extracted as well as
        the parameters defining the drift coefficient $D_1(x)$
        which controls the deterministic part of the underlying
        evolution equation (see text).}
\label{fig8}
\end{figure}

Using the same data as generated in Fig.~\ref{fig5} with $\sigma=1$, we
now plot in Fig.~\ref{fig7} the functions $\hat{m}_i$ and $\hat{\gamma}_i$ 
for the data (symbols) and compare them with the integral forms of those 
functions
for the first estimate of parameter values (dashed lines) and the optimized
solution obtained with the Levenberg-Marquardt procedure (solid lines).
Clearly, the optimized functions fit better the data and the minimum of
$F$ found is very close to its true value 
(see caption of Fig.~\ref{fig7}).

Notice that the optimized values $d_{ik}^{\prime}$ are obtained for the
transformed data ($x\to x^{\prime}=x-\langle x\rangle$), assuming
$d_{10}^{\prime}=0$. In practice one obtains $d_{10}^{\prime}\sim 10^{-2}$,
typically two orders of magnitude smaller than the other coefficients.
Using $\langle x\rangle = - d_{10}/d_{11}$, one
obtains the true coefficients according to
$d_{10}=-d_{11}^{\prime}\langle x\rangle$,
$d_{11}=d_{11}^{\prime}$,
$d_{20}=d_{20}^{\prime}-d_{21}^{\prime}\langle x\rangle +
       d_{22}^{\prime}\langle x\rangle^2$,
$d_{21}=d_{21}^{\prime}-2d_{22}^{\prime}\langle x\rangle$ and
$d_{22}=d_{22}^{\prime}$.

To show the power of the present procedure we next generate several
synthetic data sets from Eq.~(\ref{xlangevin}) with different 
measurement noise amplitudes $\sigma_I$ in the range $[0,1.2]$. 
The same $D_1(x)$ and $D_2(x)$ as in Fig.~\ref{fig2} is used.
Results are shown in Fig.~\ref{fig8}. 
The circles indicate the obtained parameter values for the first estimate, 
as in Fig.~\ref{fig2}. The solid lines indicate the true values used to
generate the data, while bullets indicate the value after optimization.

From Fig.~\ref{fig8}a one sees that after optimization the value of
$\sigma_I$ is always correctly determined.
Such finding is of major importance and shows the relevance of our
approach for practical applications even for strong measurement noise,
since the uncontaminated series $x$ typically lies within the range 
$[-2,2]$, having therefore values close to the amplitude 
$\sigma_I$ of the measurement noise.

Figures \ref{fig8}b and \ref{fig8}c also show a very reliable estimate for
the two parameters $d_{10}$ and $d_{11}$ respectively, defining the drift 
coefficient $D_1(x)$.
Since this coefficient characterizes the deterministic part of the 
evolution equation for $x$, this accurate estimate should provide valuable 
insight into the dynamics of the underlying system.

As for the diffusion coefficient $D_2(x)$, Figs.~\ref{fig8}d-f show
that the estimate of $d_{22}$ is no longer as good as for 
the other parameters.
Parameter $d_{20}$ is reasonably estimated but the optimized estimate
is as good as the first one.

For stronger measurement noise, 
namely for $\sigma>1.2$,
one faces the problem that the optimization procedure is
sometimes stucked in a local minimum of the function $F$
leading to unreliable coefficients $d_{ik}$. This is in principle
a shortcoming of the presently used minimization algorithm.
In addition, the function $F$ itself is based on estimated
functions $m$ and $\gamma$ and therefore itself subject
to errors. A forthcoming study will address the observed issues
in the context of global optimization.

\section{The North Atlantic Oscillation: an empirical example}
\label{sec:applications}

In this section we apply our framework to the North Atlantic Oscillation 
daily index, which presents data with strong measurement noise.
Table \ref{tab1} summarizes the optimized values for all 
parameter describing the data set, comparing it with simulations.
\begin{figure}[htb]
\begin{center}
\includegraphics*[width=8.5cm,angle=0]{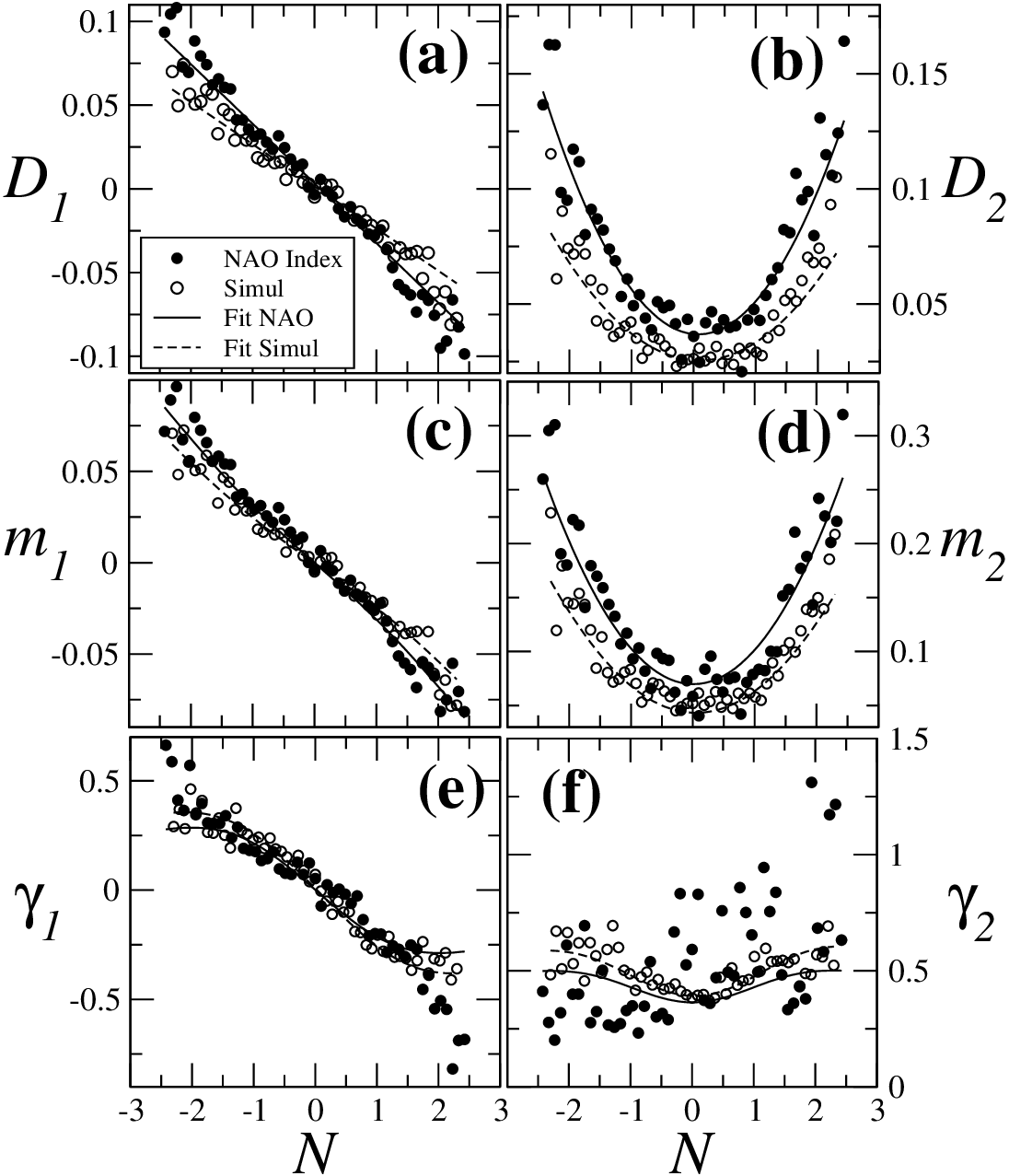}
\end{center}
\caption{\protect
     {\bf (a)-(b)} Estimate of the drift and diffusion coefficients
     $D_1(N)$ and $D_2(N)$ of the daily North Atlantic Index 
     $N$\cite{ijbc} ($16801$ datapoints), together with the corresponding
     {\bf (c)} $m_1(N)$,
     {\bf (d)} $m_2(N)$,
     {\bf (e)} $\gamma_1(N)$ and
     {\bf (f)} $\gamma_2(N)$.
     Results for the empirical NAO index are represented with bullets
     whereas the synthetic data also with $16801$ datapoints and parameter
     values given by Tab.~\ref{tab1} is shown with circles for comparison.
     The corresponding fits are given with solid and dashed lines, 
     respectively.}
\label{fig9}
\end{figure}

The North Atlantic Oscillation (NAO) is a source of variability 
in the global atmosphere, describing a large-scale vacillation in 
atmospheric mass between the anticyclone near the Azores and the 
cyclone near Iceland~\cite{hurrel95}. 
The state of the NAO is usually measured by an index $N$, defined
as the normalized pressure difference between the high and the low poles,
where the pressures are averaged over each, day, month or 
year~\cite{lind05,hurrel95}.
The NAO index and climate indices in general are receiving much attention 
due to their important role in climate change.
Lately, evidences for the stochasticity of this index have been 
shown\cite{collette04,lind05}. In this section we address the problem
of estimating its measurement noise amplitude.

Figures \ref{fig9}a and \ref{fig9}b show the drift and diffusion coefficients
respectively for the NAO daily index (bullets) and the corresponding
fit (solid line). The parameters $d_{ij}$ for both $D_1$ and $D_2$ are
given in Tab.~\ref{tab1} together with the amplitude of the measurement
noise.  
Probably due to the small amount of data points ($16\ 801$ values) one
observes large scattering of the data, particularly away from the average 
value $\langle N\rangle\sim 0$.

To evaluate the reliability of considering the NAO index a Markov
process described by Eq.~(\ref{xlangevin}) we also plot in Figs.~\ref{fig9}
the results obtained when integrating such equation (circles) using the 
coefficient values in Tab.~\ref{tab1} including the amplitude of the 
measurement noise. The same sample size was considered.
The corresponding fit is represented with a dashed line.
While the drift coefficient $D_1$ resembles the one observed for the NAO
index, there is a significant shift of the diffusion coefficient, that
only for a very narrow range around the average value is well reproduced.
Indeed, as one sees from Tab.~\ref{tab1}, the coefficient values for $D_2$ in
our simulation significantly deviate from the ones found for the
NAO series.

Further, functions $m_i$ and $\gamma_i$, plotted in Figs.~\ref{fig9}c-f, 
show also large scattering, particularly for $\gamma_2$. This feature
raises difficulties in a proper minimum search for $F$.

\begin{table}[htb]
\begin{tabular}{|c||c|c|c|c|}
\hline
 & & \multicolumn{3}{c|}{\ } 
\\
\multirow{4}{1.0cm}{Param.} & 
\multicolumn{1}{c|}{\ } & 
\multicolumn{3}{c|}{\ Simulations (16801 pts, 10 sim)\ } 
\\
 & 
\multicolumn{1}{c|}{\ NAO Index\ } & 
\multicolumn{3}{c|}{\ } 
\\\cline{3-5}
   & 
\multicolumn{1}{c|}{\ } & & &  
\\
 & 
\multicolumn{1}{c|}{\ (16801 pts)\ } &  
\multicolumn{1}{c|}{\ With noise \ } & 
\multicolumn{1}{c|}{\ No noise \ } & 
\multicolumn{1}{c|}{\ Only noise \ } 
\\
   & 
\multicolumn{1}{c|}{\ } & & &  
\\\hline
 & & & & \\
\multicolumn{1}{|c||}{$\mathbf{\sigma}$ \tiny{($\times 10^{-3}$)}} & 
\multicolumn{1}{c|}{$455$} & 
\multicolumn{1}{c|}{$455\pm 34$} &
\multicolumn{1}{c|}{$106\pm 17$} &
\multicolumn{1}{c|}{$321\pm  9$} \\
 & & & & 
\\\hline
 & & & & \\
\multicolumn{1}{|c||}{\ $\mathbf{d_{10}}$ \tiny{($\times 10^{-3}$)}\ } & 
\multicolumn{1}{c|}{$-2.6$} & 
\multicolumn{1}{c|}{$-3.1\pm 0.5$} &
\multicolumn{1}{c|}{$-3.8\pm 0.1$} &
\multicolumn{1}{c|}{$10^{-6}\pm 10^{-2}$} \\
 & & & & 
\\\hline
 & & & & \\
\multicolumn{1}{|c||}{$\mathbf{d_{11}}$ \tiny{($\times 10^{-3}$)}} & 
\multicolumn{1}{c|}{$-40$} & 
\multicolumn{1}{c|}{$-24\pm 9$} &
\multicolumn{1}{c|}{$-29\pm 2$} &
\multicolumn{1}{c|}{$0.1\pm 1$} \\
 & & & & 
\\\hline
 & & & & \\
\multicolumn{1}{|c||}{$\mathbf{d_{20}}$ \tiny{($\times 10^{-3}$)}} & 
\multicolumn{1}{c|}{$39$} & 
\multicolumn{1}{c|}{$24\pm 1$} &
\multicolumn{1}{c|}{$29\pm 0.1$} &
\multicolumn{1}{c|}{$0.1\pm 1$} \\
 & & & & 
\\\hline
 & & & & \\
\multicolumn{1}{|c||}{$\mathbf{d_{21}}$ \tiny{($\times 10^{-3}$)}} & 
\multicolumn{1}{c|}{$1.5$} & 
\multicolumn{1}{c|}{$-0.3\pm 1$} &
\multicolumn{1}{c|}{$-1.6\pm 0.2$} &
\multicolumn{1}{c|}{$0.1\pm 1$} \\
 & & & & 
\\\hline
 & & & & \\
\multicolumn{1}{|c||}{$\mathbf{d_{22}}$ \tiny{($\times 10^{-3}$)}} & 
\multicolumn{1}{c|}{$16$} & 
\multicolumn{1}{c|}{$13\pm 3$} &
\multicolumn{1}{c|}{$11\pm 0.5$} &
\multicolumn{1}{c|}{$-1\pm 7$} \\
 & & & & 
\\\hline
\end{tabular}
\caption{\label{tab1}\protect
     Optimized parameter values for the daily North Atlantic 
     Oscillation daily index\cite{ijbc} compared with the average values
     for $10$ sets of synthetic data (``With noise'') using the same number 
     of points and parameter values.  
     In order to evaluate the reliability of our synthetic data we also
     run the optimization procedure for $10$ sets of synthetic data with
     the same $D_1$ and $D_2$ found in NAO series and $\sigma=0$ 
     (``No noise''). In the last column we plot the results returned from
     the optimization procedure for synthetic data of pure measurement
     noise with amplitude $\sigma=0.455$, the one obtained for NAO series.}
\end{table}

In order to check the reliability of the calculations
we reproduce the synthetic data $10$ times and present in Tab.~\ref{tab1}
(column ``With noise'')
the average values for each parameter, where the error is taken as the largest
deviation from the average over the sample of data sets.
The measurement noise, which dominates all parameters, is well 
reproduced. For the drift and diffusion coefficient the order of
magnitude of each parameter is also correct, but for $d_ {11}$ and
and $d_{20}$ one observes significant deviations from the
estimated values 
obtained for the NAO series.

This mismatch between the empirical and synthetic series could raise
the question if the NAO Index is indeed suitably described by a
Markovian stochastic process with a perceivable deterministic part.
In fact, since one observes $\sigma \gg d_{ij}$ the 
series is approximately a pure white noise (i.e.~$y(t)=\sigma \zeta(t)$
in Eq.~(\ref{yy})),
which in fact also yields a linear drift and quadratic 
diffusion coefficients.

To address this problem we rerun our optimization procedure for synthetic
data, for two additional situations, one where $\sigma=0$ and drift and
diffusion coefficients are given by the NAO index, and another one which 
simulates a pure white noise ($D_1=D_2=0$) with $\sigma$ equal to the value
found for the NAO series. The results are also given in Tab.~\ref{tab1},
columns ``No Noise'' and ``Only noise'' respectively.

For the pure white noise process one obtains $\sigma$ as the only non-zero
parameter, apart fluctuations, but with an amplitude different from the
one used to generate the synthetic data, namely $\sigma\simeq 0.321$, which
corresponds to $\sim 75\%$ of the inserted
measurement noise ($\sigma=0.445$).
For the synthetic process with no noise, the order of magnitude for the
parameters of $D_1$ and $D_2$ is correctly computed, whereas a non-zero 
measurement noise is retrieved
covering the remaining $25\%$ of the inserted measurement noise.
In other words, one can argue that in this situation our procedure 
retrieves $\sim 75\%$ of the total amount of measurement noise.

In this scope, our results point in the direction of previous arguments 
given by some authors\cite{stephenson00}: 
differently from other climate indices such as 
the ENSO index, the NAO index seems to be an almost 
pure white noise process with only a minor contribution from a 
stochastic process governed by a Langevin-like equation.
Alternative indices should be therefore considered and studied as 
recently suggested\cite{lind05}.

\section{Discussion and Conclusions}
\label{sec:conc}

We described in detail a nonparametric procedure to extract measurement 
noise in empirical stochastic series with strong measurement noise.
The algorithm is able to accurately extract the strength of measurement 
noise and the values of the parameters defining the drift coefficient and 
to estimate with good accuracy the diffusion coefficient that 
fully describe the evolution equation for the measured quantity 
in the time series. 
This has been shown by synthetically generated
data sets contaminated by increasing measurement noise. 
Additionally, the algorithm was applied to 
a set of measured data 
providing new insight in the underlying systems.
The data for the climate index shows a large scattering, probably
due to the small amount of data points. Larger data sets for climate
indices are not available up to our knowledge.

It should be noticed that the nonparametric reconstruction of the
Langevin Eq.~(\ref{xlangevin}) from measured stationary data sets
generally requires that the process exhibits Markovian properties and 
fulfils the Pawula theorem\cite{lind05}. 
While the second constraint can be relaxed extending the analysis to a 
broader class of Langevin-like systems in which the Gaussian 
$\delta$-correlated white noise 
Langevin force is replaced by a more general L\'{e}vy 
noise\cite{friedrich08,siegert01}, in general the Markov condition remains 
a crucial constraint. 

Recently, it has been shown that processes corrupted from measurement 
noise may loose their Markov properties\cite{kleinhans07}. 
For this reason the proper 
analysis of data suffering from strong measurement noise in general is a 
complicated task. We, however, would like to point out, that the method 
presented here solely relies on Markov properties of the underlying, 
undisturbed process $x(t)$. In case of $\delta$-correlated measurement 
noise the method presents a general approach to access the process $x$ 
and the noise amplitude $\sigma$ at the same time.

Therefore, since the algorithm is general for a broad class of 
stochastic systems other applications can be proposed.
Particularly in cases where the measurement procedure is subject
to large measurement noise due to the distance between the location
where the measure is taken and the location where the phenomena 
occurs. Two important applications in this context are seismographic
data\cite{friedrich08}, where the epicenter can not be predicted before-hand,
and data from surface EEG\cite{prusseit08,lehnertz09}, which, though having 
stronger measurement noise, are much recommended instead of {\it insitu}
measurements for the sake and comfort of the patient.
A further application would be the analysis of sensors to which one has
no access, for example sensors being installed in remote systems
showing more and more measurement noise due to aging effects.
Here it 
should even be possible to know quite precisely the functional structure of
the underlying process, an assumption of our analysis here.

Such applications however appeal for the extension of the present
procedures to higher dimensions, i.e.~more than one time-series, which
implies the consideration of different measurement noise sources and
consequently noise mixing. To ascertain in which conditions and up to
which point can we separate different measurement noise sources is 
an open question which we will address elsewhere.

In all simulations a linear function was assumed for the drift coefficient
and a quadratic one for diffusion. 
Although such assumptions comprehend already a broad class of 
systems\cite{friedrich08,lind05,boettcher06} 
our approach and all expressions may easily be extended to higher
order polynomials for $D_1(x)$ and $D_2(x)$, as long as the number of 
parameters for modelling $D_1(x)$ and $D_2(x)$ is not too high.
In this case the calculations presented in the appendices are valid
if one considers proper higher powers in the integrand of integrals
$h_1$ and $h_2$ (see Eqs.~(\ref{hmore}) in Append.~\ref{app:deriv}).

Furthermore, other possibilities for optimization are possible.
For instance, though in this case we have shown that random Monte Carlo 
procedures are computationally expensive consuming, one could think
of a non-local search procedure using for example bigger jumps such as 
the ones of a L\'evy flight process\cite{pavlyukevich07}.
Alternatively one may also study how good would be an optimization
procedure that considers the minimization of a splitted cost function
$F$.
Preliminary results have shown that for a proper decomposition of
$F$ our optimization problem may be reduced to a cubic equation
and a lower dimensional system of linear equations.
Another possibility would be to use genetic algorithms\cite{genetic}.
These points will be addressed elsewhere.

\section*{Acknowledgements}

The authors thank Wilhelm and Else Heraeus Foundation for supporting the
meeting hold in Bad Honnef, where very usefull discussions happened and
also the project DREBM/DAAD/03/2009 for the bilateral cooperation between
Portugal and Germany. PGL thanks Reza M.~Baram and Bibhu Biswal for 
usefull discussions.

\appendix
\begin{widetext}
\section{The conditional moments of an arbitrary time series
              and their linear approximations}
\label{app:km}

Taking a series of measurements $y(t)$ as defined in Eq.~(\ref{yy}),
its $n$-th order conditional moment reads
\begin{eqnarray}
\hat{M}_n(y_0,\tau) &=& \langle (y(t+\tau)-y(t))^n \rangle |_{y(t)=y_0}
\cr
   & & \cr
   &=& \int_{-\infty}^{+\infty} dx_0 \int_{-\infty}^{+\infty} dx
       \int_{-\infty}^{+\infty} dy (y-y_0)^n f_{\sigma}(y\vert x) 
       f_{\tau}(x\vert x_0)\bar{f}_{\sigma}(x_0\vert y_0),
\label{gencondmom_app}  
\end{eqnarray}
where $f_{\sigma}(y\vert x)$ is the probability to measure $y$ in 
the presence of a measurement noise with variance $\sigma^2$, when the 
system (without noise) has the value $x$,
$f_{\tau}(x\vert x_0)$ is the probability for the system to evolve from 
a value $x_0$ to a value $x$ within a time interval
$\tau$ and
$\bar{f}_{\sigma}(x_0\vert y_0)$ has the inverse meaning of $f_{\sigma}$:
it is the probability for the system to adopt the value $x_0$ when a 
measured value $y_0$ is observed.
While $f_\tau$ is unknown, $f_{\sigma}$ and $\bar{f}_{\sigma}$ are related 
with each other according to Bayes' theorem (see App.~\ref{app:condprob}).

From such assumptions one easily arrives to the identities
\begin{subequations}
\begin{eqnarray}
\int_{-\infty}^{+\infty} dy f_{\sigma}(y\vert x) &=& 1 ,\label{eq1}\\
\int_{-\infty}^{+\infty} dy (y-x) f_{\sigma}(y\vert x) &=& 0 ,\label{eq2}\\
\int_{-\infty}^{+\infty} dy (y-x)^2 f_{\sigma}(y\vert x) &=& \sigma^2 ,\label{eq3}
\end{eqnarray}\label{eq}\end{subequations}
and using these identities the general expression (\ref{gencondmom_app})
can be approximated up to first order assuming $\tau\ll 1$.
More precisely, the first two moments $\hat{M}_1$ and $\hat{M}_2$ yield
\begin{eqnarray}
\hat{M}_1(y_0,\tau) &=& \langle y(t+\tau)-y(t) \rangle |_{y(t)=y_0}\cr
   & & \cr
   &=& \int_{-\infty}^{+\infty} dx_0 \int_{-\infty}^{+\infty} dx
       \int_{-\infty}^{+\infty} dy (y-y_0) f_{\sigma}(y\vert x) 
       f_{\tau}(x\vert x_0)\bar{f}_{\sigma}(x_0\vert y_0), \cr
   & & \cr
   &=& \int_{-\infty}^{+\infty} dx_0 
       \int_{-\infty}^{+\infty} dx   f_{\tau}(x\vert x_0)
                                     \bar{f}_{\sigma}(x_0\vert y_0)
       \int_{-\infty}^{+\infty} dy (y-x+x-y_0) f_{\sigma}(y\vert x)  \cr
   & & \cr
   &=& \int_{-\infty}^{+\infty} dx_0 
       \int_{-\infty}^{+\infty} dx   f_{\tau}(x\vert x_0)
                                     \bar{f}_{\sigma}(x_0\vert y_0)\times\cr
   & & \cr
   & & \times
       \left (
       \int_{-\infty}^{+\infty} dy (x-y_0) f_{\sigma}(y\vert x) 
      +\int_{-\infty}^{+\infty} dy (y-x) f_{\sigma}(y\vert x) 
       \right ) \cr
   & & \cr
   &=& \int_{-\infty}^{+\infty} dx_0 
       \int_{-\infty}^{+\infty} dx   f_{\tau}(x\vert x_0)
                                     \bar{f}_{\sigma}(x_0\vert y_0)\times
       \left (
       (x-y_0)\int_{-\infty}^{+\infty} dy f_{\sigma}(y\vert x) 
      + 0 
       \right ) \cr
   & & \cr
   &=& \int_{-\infty}^{+\infty} dx_0 \bar{f}_{\sigma}(x_0\vert y_0)
       \int_{-\infty}^{+\infty} dx (x-x_0+x_0-y_0)
       f_{\tau}(x\vert x_0) \cr
   & & \cr
   &=& \int_{-\infty}^{+\infty} dx_0 \bar{f}_{\sigma}(x_0\vert y_0)\times\cr
   & & \cr
   & & \times    \left (
       \int_{-\infty}^{+\infty} dx (x_0-y_0) f_{\tau}(x\vert x_0) +
       \int_{-\infty}^{+\infty} dx (x-x_0) f_{\tau}(x\vert x_0) 
       \right )\cr
   & & \cr
   &=& \int_{-\infty}^{+\infty} dx_0 \bar{f}_{\sigma}(x_0\vert y_0)
       \left (
       (x_0-y_0) +
       \tau D_1(x_0) + {\cal O}(\tau^2) 
       \right ) \cr
   & & \cr
   &=& \int_{-\infty}^{+\infty} dx_0 (x_0-y_0)\bar{f}_{\sigma}(x_0\vert y_0)
      +\tau\int_{-\infty}^{+\infty} dx_0 D_1(x_0)\bar{f}_{\sigma}(x_0\vert y_0)
      +{\cal O}(\tau^2) \cr
   & & \cr
   &\equiv& \hat{\gamma}_1(y_0) + \tau \hat{m}_1(y_0) + {\cal O}(\tau^2), 
            \label{yM1_2_app}\\
   & & \cr
\hat{M}_2(y_0,\tau) &=& \langle (y(t+\tau)-y(t))^2 \rangle |_{y(t)=y_0}\cr
   & & \cr
   &=& \int_{-\infty}^{+\infty} dx_0 \int_{-\infty}^{+\infty} dx
       \int_{-\infty}^{+\infty} dy (y-y_0)^2 f_{\sigma}(y\vert x) 
       f_{\tau}(x\vert x_0)\bar{f}_{\sigma}(x_0\vert y_0), \cr
   & & \cr
   &=& \int_{-\infty}^{+\infty} dx_0 \int_{-\infty}^{+\infty} dx
       f_{\tau}(x\vert x_0)\bar{f}_{\sigma}(x_0\vert y_0)
       \int_{-\infty}^{+\infty} dy (y-y_0)^2 f_{\sigma}(y\vert x) \cr
   & & \cr
   &=& \int_{-\infty}^{+\infty} dx_0 \int_{-\infty}^{+\infty} dx
       f_{\tau}(x\vert x_0)\bar{f}_{\sigma}(x_0\vert y_0)
       \int_{-\infty}^{+\infty} dy (y-x+x-y_0)^2 f_{\sigma}(y\vert x) \cr
   & & \cr
   &=& \int_{-\infty}^{+\infty} dx_0 \int_{-\infty}^{+\infty} dx
       f_{\tau}(x\vert x_0)\bar{f}_{\sigma}(x_0\vert y_0) \times\cr
   & & \cr
   & & \times
       \Big (
       \int_{-\infty}^{+\infty} dy (y-x)^2 f_{\sigma}(y\vert x) +
       2(x-y_0)\int_{-\infty}^{+\infty} dy (y-x) f_{\sigma}(y\vert x) +\cr
   & & \cr
   & &  \hspace{0.2cm}
       (x-y_0)^2 \int_{-\infty}^{+\infty} dy f_{\sigma}(y\vert x) 
       \Big ) \cr
   &=& \int_{-\infty}^{+\infty} dx_0 \int_{-\infty}^{+\infty} dx
       f_{\tau}(x\vert x_0)\bar{f}_{\sigma}(x_0\vert y_0) \times\
       \left ( \sigma^2 + 0 + (x-y_0)^2 \right ) \cr
   & & \cr
   &=& \int_{-\infty}^{+\infty} dx_0 \bar{f}_{\sigma}(x_0\vert y_0) 
       \int_{-\infty}^{+\infty} dx
        (\sigma^2 + (x-y_0)^2 ) 
       f_{\tau}(x\vert x_0) \cr
   & & \cr
   &=& \int_{-\infty}^{+\infty} dx_0 \bar{f}_{\sigma}(x_0\vert y_0) 
       \int_{-\infty}^{+\infty} dx
        (\sigma^2 + (x-x_0+x_0-y_0)^2 ) 
       f_{\tau}(x\vert x_0) \cr
   & & \cr
   &=& \int_{-\infty}^{+\infty} dx_0 \bar{f}_{\sigma}(x_0\vert y_0) \times\cr
   & & \cr
   & & \times
       \Big (
       \int_{-\infty}^{+\infty} dx
       (x-x_0)^2 f_{\tau}(x\vert x_0) +
       2(x_0-y_0)\int_{-\infty}^{+\infty} dx
       (x-x_0) f_{\tau}(x\vert x_0) \cr
   & & \cr
   & & \vspace{0.2cm}
       (\sigma^2+(x_0-y_0)^2)\int_{-\infty}^{+\infty} dx
       f_{\tau}(x\vert x_0) \Big )\cr
   &=& \int_{-\infty}^{+\infty} dx_0 
       \Big (
       2\tau D_2(x_0) + 2(x_0-y_0)\tau D_1(x_0) + \cr
   & & \cr
   & & \phantom{aaaaaaaa} \sigma^2+(x_0-y_0)^2 
       \Big ) \bar{f}_{\sigma}(x_0\vert y_0) + {\cal O}(\tau^2)\cr
   & & \cr
   &=& 2\tau\int_{-\infty}^{+\infty} dx_0 
       \Big ( D_2(x_0) + (x_0-y_0) D_1(x_0) \Big )
       \bar{f}_{\sigma}(x_0\vert y_0) + \cr 
   & & \cr
   & & \sigma^2 + \int_{-\infty}^{+\infty} dx_0 
       (x_0-y_0)^2 \bar{f}_{\sigma}(x_0\vert y_0) + {\cal O}(\tau^2)\cr
   & & \cr
   &\equiv& \tau \hat{m}_2(y_0) + \sigma^2 + 
            \hat{\gamma}_2(y_0) + {\cal O}(\tau^2). 
\label{yM2_2_app}
\end{eqnarray}

From Eq.~(\ref{yM2_2_app}) one has 
$\hat{M}(y_0,0)=\sigma^2+\hat{\gamma}_2(y_0)$
where $\hat{\gamma}_2(y_0) = \int_{-\infty}^{+\infty} dx_0 
(x_0-y_0)^2 \bar{f}_{\sigma}(x_0\vert y_0)$.
Such observations justify the first estimate for the measurement noise
stated in Eq.~(\ref{approxmeasnoise}), since
when $\sigma$ is small enough, probability density function 
$\bar{f}_{\sigma}(x_0\vert y_0)$ is similar to
$f_{\sigma}(y_0\vert x_0)$ (see Eq.~(\ref{eq1})) and therefore, one
can take as a first approximation $\bar{\gamma}_2(y_0)\sim \sigma^2$.

Notice that the last equalities in $\hat{M}_1$ and $\hat{M}_2$ yield
first order approximations under the assumption that $\tau \ll 1$.
In Ref.~\cite{kleinhans05} another approach is proposed
for the estimation of drift and diffusion coefficients
in the case of low sampling rates.

The errors for $\hat{\gamma}_1(y_0)$, $\hat{\gamma}_2(y_0)$,
$\hat{m}_1(y_0)$  and $\hat{m}_2(y_0)$ are just given from the
linear fit of $\hat{M}_1$ and $\hat{M}_2$ for each fixed $y_0$,
given in Eqs.~(\ref{yM1_2}) and (\ref{yM2_2}).
The errors of $\hat{M}_1(y,\tau)$ and $\hat{M}_2(y,\tau)$ can be
also directly computed from the data as
\begin{subequations}
\begin{eqnarray}
\sigma_{\hat{M}_1}^2(y,\tau)&=&
        \langle \left [ (y(t+\tau)-y(t))-
                        \langle y(t+\tau)-y(t)\rangle
        \right ]^2  \rangle_{t\in \{ t_1,\dots,t_n\}}\cr
 & & \cr
 &=& \langle (y(t+\tau)-y(t))^2+ \hat{M}_1^2(y_0,\tau) - 2\hat{M}_1(y_0,\tau)
     (y(t+\tau)-y(t))\rangle \cr
 & & \cr
 &=& \tfrac{1}{N_y}
     \left (
     \hat{M}_2(y_0,\tau) + \hat{M}_1^2(y_0,\tau) -2\hat{M}_1^2(y_0,\tau)
     \right )\cr
 & & \cr
 &=& \frac{\hat{M}_2(y,\tau)-\hat{M}_1^2(y,\tau)}
                           {N_y}\label{errorM1_app}\\
  & & \cr
\sigma_{\hat{M}_2}^2(y,\tau)&=&
        \langle \left [ (y(t+\tau)-y(t))^2-
                        \langle ( y(t+\tau)-y(t))^2\rangle
        \right ]^2  \rangle_{t\in \{ t_1,\dots,t_n\}}\cr
 & & \cr
 &=& \langle (y(t+\tau)-y(t))^4+ \hat{M}_2^2(y_0,\tau) - 2\hat{M}_2(y_0,\tau)
     (y(t+\tau)-y(t))^2\rangle \cr
 & & \cr
 &=& \tfrac{1}{N_y}
     \left (
     \hat{M}_4(y_0,\tau) + \hat{M}_2^2(y_0,\tau) -2\hat{M}_2^2(y_0,\tau)
     \right )\cr
 & & \cr
 &=& \frac{\hat{M}_4(y,\tau)-\hat{M}_2^2(y,\tau)}{N_y}\label{errorM2_app} ,
\end{eqnarray}
\label{errors_app}
\end{subequations}
where $N_y$ is the number of data points in bin $y$.

For the optimization procedure it is convenient to simplify the
expressions for functions $m_i$ and $\gamma_i$ ($i=1,2$). Namely,
$m_1$ and $m_2$ can be written as expressions of $\gamma_1$ and $\gamma_2$.
In fact, substituting Eqs.~(\ref{D1}) and (\ref{D2}) into Eqs.~(\ref{m1})
and (\ref{m2}), and adding and subtracting properly $y$, yields 
\begin{subequations}
\begin{eqnarray}
m_1(y) &=& \int_{-\infty}^{+\infty} D_1(x) \bar{f}_{\sigma}(x\vert y)dx \cr
       & & \cr
       &=& \int_{-\infty}^{+\infty} (d_{10}+d_{11}x) 
           \bar{f}_{\sigma}(x\vert y)dx \cr
       & & \cr
       &=& \int_{-\infty}^{+\infty} [d_{10}+d_{11}(x+y-y)] 
           \bar{f}_{\sigma}(x\vert y)dx \cr
       & & \cr
       &=& d_{10}\int_{-\infty}^{+\infty} 
           \bar{f}_{\sigma}(x\vert y)dx + 
           d_{11}\int_{-\infty}^{+\infty} (x-y)\bar{f}_{\sigma}(x\vert y)dx + 
           d_{11}y\int_{-\infty}^{+\infty} \bar{f}_{\sigma}(x\vert y)dx \cr
       & & \cr
       &=& d_{10} + d_{11}(y+\gamma_1(y)) \label{newm1}\\
       & & \cr
m_2(y) &=& 2\int_{-\infty}^{+\infty} [(x-y)D_1(x)+D_2(x)] 
           \bar{f}_{\sigma}(x\vert y)dx \cr
       & & \cr
       &=& 2\int_{-\infty}^{+\infty} 
           [(x-y)(d_{10}+d_{11}x)+d_{20}+d_{21}x+d_{22}x^2]
           \bar{f}_{\sigma}(x\vert y)dx \cr
       & & \cr
       &=& 2\int_{-\infty}^{+\infty} \Big [
                (x-y)\left ( d_{10}+d_{11}(x-y+y)\right )+
                d_{20}+ \cr
       & & \cr
       & & \hspace{2.0cm} d_{21}(x-y+y)+d_{22}(x-y+y)^2 \Big ]
                \bar{f}_{\sigma}(x\vert y)dx \cr
       & & \cr
       &=& 2 \big [ \gamma_1(y) d_{10} +
           (\gamma_2(y)+y\gamma_1(y)) d_{11} +
           d_{20} +
           (\gamma_1(y) + y ) d_{21} +
           ( 2y\gamma_1(y) + \gamma_2(y) + y^2 ) d_{22} 
           \big ] .\label{newm2}
\end{eqnarray}
\label{newm}
\end{subequations}

Substituting Eqs.~(\ref{newm1}) and (\ref{newm2}) into Eq.~(\ref{F})
yields $F$ as a functional depending only on the integrals $\gamma_1(y)$
and $\gamma_2(y)$ defined in Eqs.~(\ref{gamma1}) and (\ref{gamma2}),
apart the six parameters, $\sigma$ and $d_{jk}$, we want to optimize.

\section{The probability density function $\bar{f}_{\sigma}(x\vert y)$}
\label{app:condprob}

To solve the minimization problem we will need to explicitly write
expressions for $\bar{f}_{\sigma}(x\vert y)$. 
This conditional probability density function appears in Eqs.~(\ref{gamma1}) 
and (\ref{gamma2}) and according to the Bayes theorem is given by:
\begin{equation}
\bar{f}_{\sigma}(x\vert y) = 
\frac{f_{\sigma}(y\vert x)p(x)}{\int_{-\infty}^{+\infty} f_{\sigma}(y\vert x^{\prime})p(x^{\prime})dx^{\prime}}
\label{fx|y}
\end{equation}
where $f_{\sigma}(y\vert x)$ is the probability density function of the
measurement noise $\sigma\zeta_t$, i.e.~a Gaussian function
centered at $y$ with variance $\sigma^2$, 
\begin{equation}
f_{\sigma}(y\vert x) = \frac{1}{\sigma\sqrt{2\pi}}\hbox{\Large{e}}^{-\frac{(y-x)^2}{2\sigma^2}} ,
\label{noisefunc}
\end{equation}
and $p(x)$ can be written, assuming that the process is stationary, as
\begin{equation}
p(x) = \frac{{\cal N}}{D_2(x)}
       \hbox{\Large{e}}^{\Phi(x)}
\label{px}
\end{equation}
where ${\cal N}$ is some normalized function such that
$\int_{-\infty}^{\infty}p(x)dx=1$ and 
\begin{equation}
\Phi(x)=\int_{-\infty}^x\frac{D_1(x^{\prime})}{D_2(x^{\prime})}dx^{\prime} .
\label{phifunc}
\end{equation}

For an Ornstein-Uhlenbeck process $D_1(x)=d_{10}+d_{11}x$ and
$D_2(x)=d_{20}$ one finds
\begin{equation}
p_{OU}(x)=\sqrt{-\frac{d_{11}}{2d_{20}\pi}}
          \hbox{\Large{e}}^{\frac{1}{2}\frac{d_{11}}{d_{20}}(x+\frac{d_{10}}{d_{11}})^2} ,
\end{equation}
from which one easily sees that $d_{11}<0$ is a necessary condition to
have a well-defined probability density function $p(x)$.

For the general case given by Eqs.~(\ref{DD}) one has typically
$D_2(x)>0$ with $d_{22}>0$, which yields $\Delta\equiv
4d_{20}d_{22}-d_{21}^2>0$.
In these situations, $p(x)$ can also be integrated, yielding
\begin{equation}
p_G(x) = {\cal N} 
       (D_2(x))^{\frac{d_{11}}{2d_{22}}-1}
       \hbox{\Large{e}}^{(d_{10}-\frac{d_{21}d_{11}}{2d_{22}})h_0(x)} , 
\label{genpx}
\end{equation}
with
\begin{equation}
h_0(x) = \frac{2}{\sqrt{\Delta}}
         \left [
         \arctan{\left (
                 \frac{2d_{22}x+d_{21}}{\sqrt{\Delta}}
                 \right )} + \tfrac{\pi}{2} \right ] .
\label{h0}
\end{equation}

\section{The derivatives of 
              $\gamma_1$, $\gamma_2$, $m_1$ and $m_2$}
\label{app:deriv}

The minimization problem needs also the expression of the derivatives
for the $\gamma$'s and $m$'s.
To compute them one needs first to write the derivatives of 
function $\bar{f}_{\sigma}(x\vert y)$ defined in Eq.~(\ref{fx|y}).

Defining $g(x,y)\equiv f_{\sigma}(y\vert x)p(x)$ one has in general
\begin{equation}
\frac{\partial \bar{f}_{\sigma}(x\vert y)}{\partial X} =
\frac{\frac{\partial g}{\partial X}\int_{-\infty}^{+\infty} g(x^{\prime},y)dx^{\prime} -
      g \int_{-\infty}^{+\infty} \frac{\partial g}{\partial X}dx^{\prime}}
        {\left ( \int_{-\infty}^{+\infty} g(x^{\prime},y)dx^{\prime}\right )^2},
\end{equation}
where $X$ is some variable on which $\bar{f}_{\sigma}$ depends.
Since $p(x)$ depends only on parameters $d_{ij}$ and 
$f_{\sigma}(y\vert x)$ depends only on $\sigma$, we have
\begin{subequations}
\begin{eqnarray}
\frac{\partial g(x,y)}{\partial \sigma} &=&
        \frac{\partial f_{\sigma}(y\vert x)}{\partial \sigma} p(x) \label{dgsigma}\\ 
\frac{\partial g(x,y)}{\partial d_{ij}} &=&
        \frac{\partial p(x)}{\partial d_{ij}} f_{\sigma}(y\vert x) \label{dgds}
\end{eqnarray}
\end{subequations}
where for $f_{\sigma}(y\vert x)$ we have
\begin{equation}
\frac{\partial f_{\sigma}(y\vert x)}{\partial \sigma} =
        f_{\sigma}(y\vert x) \frac{(x-y)^2}{\sigma^3} \label{dfdsig}
\end{equation}
and for $p(x)$ we have
\begin{equation}
p(x)=\frac{{\cal N}}{D_2(x)}\hbox{\Large e}^{\Phi(x)}\equiv {\cal N}\hat{p}(x) ,
\end{equation}
with
\begin{equation}
{\cal N} = \left ( 
           \int_{-\infty}^{+\infty} \hat{p}(x) dx
           \right )^{-1}
\end{equation}
and therefore
\begin{equation}
\frac{\partial p(x)}{\partial X} =
   {\cal N} \left (
            \frac{\partial \hat{p}(x)}{\partial X} 
    - p(x) \int_{-\infty}^{+\infty} \frac{\partial \hat{p}(x^{\prime})}{\partial X} dx^{\prime}
            \right ) 
\end{equation}
with $X$ one of the $d$ parameters.

In the Ornstein-Uhlenbeck case
\begin{subequations}
\begin{eqnarray}
\frac{\partial \hat{p}_{OU}(x)}{\partial d_{10}} &=&
        \frac{1}{d_{20}}(x+\frac{d_{10}}{d_{11}})p_{OU}(x) \label{dpdd10OU}\\
\frac{\partial \hat{p}_{OU}(x)}{\partial d_{11}} &=&
        \left (
        \frac{1}{2d_{20}}(x^2-\frac{d_{10}^2}{d_{11}^2})
        +\frac{1}{2d_{11}}
        \right ) p_{OU}(x) 
                \label{dpdd11OU}\\
\frac{\partial \hat{p}_{OU}(x)}{\partial d_{20}} &=&
        -\frac{1}{2d_{20}} \left (
         1+\frac{d_{11}}{d_{20}}
            \left (
               x+\frac{d_{10}}{d_{11}}
            \right )^2
         \right ) p_{OU}(x) \label{dpdd20OU}
\end{eqnarray}
\end{subequations}
and in the general case
\begin{subequations}
\begin{eqnarray}
\frac{\partial \hat{p}_G(x)}{\partial d_{10}} &=& h_0(x) p_G(x) \label{dpdd10}\\
\frac{\partial \hat{p}_G(x)}{\partial d_{11}} &=&
        \left (
           \frac{1}{2d_{22}}\log{D_2(x)}-\frac{d_{21}}{2d_{22}}h_0(x)
        \right ) p_G(x) \label{dpdd11}\\
\frac{\partial \hat{p}_G(x)}{\partial d_{20}} &=&
        \left (
            \frac{1}{D_2(x)} + \frac{\partial h_0(x)}{\partial d_{20}}
         \right ) p_G(x) \label{dpdd20}\\
\frac{\partial \hat{p}_G(x)}{\partial d_{21}} &=&
        \left (
            \frac{x}{D_2(x)} - \frac{d_{11}}{2d_{22}}h_0(x)+
            \left (
              d_{10}-\frac{d_{21}d_{11}}{2d_{22}}
            \right )
            \frac{\partial h_0(x)}{\partial d_{21}}
         \right ) p_G(x) \label{dpdd21}\\
\frac{\partial \hat{p}_G(x)}{\partial d_{22}} &=&
        \left (
            \left( 
               \frac{d_{11}}{2d_{22}}-1 
            \right )
            \frac{x^2}{D_2(x)} - 
            \frac{d_{11}}{2d_{22}^2}\log{D_2(x)}+
            \frac{d_{21}d_{11}}{2d_{22}^2}h_0(x)+
            \left (
              d_{10}-\frac{d_{21}d_{11}}{2d_{22}}
            \right )
            \frac{\partial h_0(x)}{\partial d_{22}}
         \right ) p_G(x) \label{dpdd22}
\end{eqnarray}
\end{subequations}
where
\begin{subequations}
\begin{eqnarray}
\frac{\partial h_0(x)}{\partial d_{20}}  &=&  
              -\frac{2d_{22}}{\Delta}h_0(x) -
               \frac{4d_{22}(2d_{22}x+d_{21})}
                 {\Delta (\Delta + (2d_{22}x+d_{21})^2)}    \label{dh0d20}\\
\frac{\partial h_0(x)}{\partial d_{21}}  &=& 
               \frac{d_{21}}{\Delta}h_0(x)   +   
               \frac{2}{\Delta} 
               \frac{\Delta + d_{21}(2d_{22}x+d_{21})}
                 {\Delta + (2d_{22}x+d_{21})^2}    \label{dh0d21}\\
\frac{\partial h_0(x)}{\partial d_{22}}  &=&  
               -\frac{2d_{20}}{\Delta}h_0(x)   +   
               \frac{4}{\Delta} 
               \frac{x\Delta - d_{20}(2d_{22}x+d_{21})}
                 {\Delta + (2d_{22}x+d_{21})^2}     .\label{dh0d22}
\end{eqnarray}
\end{subequations}

So, neglecting the parameter $d_{10}$ as explained in Sec.~\ref{sec:algor},
for the other parameters $\sigma,d_{11},d_{20},d_{21},d_{22}$ we have
\begin{subequations}
\begin{eqnarray}
\frac{\partial \bar{f}_{\sigma}(x\vert y)}{\partial \sigma} &=&
      \frac{1}{\sigma^3} \left [
            (x-y)^2-\gamma_2(y)
                         \right ] \bar{f}_{\sigma}(x\vert y) 
         \label{der-f-sigma}\\
     & & \cr
\frac{\partial \bar{f}_{\sigma}(x\vert y)}{\partial d_{ij}} &=&
      \frac{\hbox{\Large{e}}^{-\frac{(x-y)^2}{2\sigma^2}}
            \frac{\partial p(x)}{\partial d_{ij}}
            -\bar{f}_{\sigma}(x\vert y)
            \int_{-\infty}^{+\infty} 
            \hbox{\Large{e}}^{-\frac{(x^{\prime}-y)^2}{2\sigma^2}}
            \frac{\partial p(x^{\prime})}{\partial d_{ij}}dx^{\prime}}
           {\int_{-\infty}^{+\infty} g(x^{\prime},y)dx^{\prime}} \label{der-f-dij}
\end{eqnarray}
\label{der-f}
\end{subequations}
and therefore considering Eqs.~(\ref{gamma1}) and (\ref{gamma2})
that define functions $\gamma_1(y)$ and $\gamma_2(y)$ and also
Eqs.~(\ref{newm1}) and (\ref{newm2}) defining functions
$m_1(y)$ and $m_2(y)$ it follows
\begin{subequations}
\begin{eqnarray}
\frac{\partial \gamma_1(y)}{\partial \sigma} &=&
      \frac{1}{\sigma^3} \left [ h_1(y) - \gamma_1(y)\gamma_2(y)
                         \right ] \label{der-gamma1-sigma}\\
     & & \cr
\frac{\partial \gamma_2(y)}{\partial \sigma} &=&
      \frac{1}{\sigma^3} \left [ h_2(y) - \gamma_2^2(y)
                         \right ] \label{der-gamma2-sigma}\\
     & & \cr
\frac{\partial m_1(y)}{\partial \sigma} &=& d_{11} 
           \frac{\partial \gamma_1(y)}{\partial \sigma} 
   \label{der-m1-sigma}\\
   & & \cr
\frac{\partial m_2(y)}{\partial \sigma} &=& 
     2\left (
        (d_{21}+y(d_{11}+2d_{22}))
     \frac{\partial \gamma_1(y)}{\partial \sigma} 
        + (d_{11}+d_{22})
     \frac{\partial \gamma_2(y)}{\partial \sigma}  
     \right )
   \label{der-m2-sigma}\\
   & & \cr
\frac{\partial \gamma_1(y)}{\partial d_{ij}} &=&
      \int_{-\infty}^{+\infty} (x^{\prime}-y) 
        \frac{\partial \bar{f}_{\sigma}(x^{\prime}\vert y)}{\partial d_{ij}} dx^{\prime} 
    \label{der_gamma1_dij}\\
    & & \cr
\frac{\partial \gamma_2(y)}{\partial d_{ij}} &=&
      \int_{-\infty}^{+\infty} (x^{\prime}-y)^2 
        \frac{\partial \bar{f}_{\sigma}(x^{\prime}\vert y)}{\partial d_{ij}} dx^{\prime} 
    \label{der_gamma2_dij}\\
    & & \cr
\frac{\partial m_1(y)}{\partial d_{11}} &=& 
           y+\gamma_1(y) + d_{11} 
           \frac{\partial \gamma_1(y)}{\partial d_{11}} 
   \label{der-m1-d11}\\
   & & \cr
\frac{\partial m_1(y)}{\partial d_{2j}} &=& 
           d_{11} \frac{\partial \gamma_1(y)}{\partial d_{2j}} 
   \label{der-m1-d2j}\\
   & & \cr
\frac{\partial m_2(y)}{\partial d_{11}} &=& 
           2\left (
             (d_{21}+y(d_{11}+2d_{22}))
             \frac{\partial \gamma_1(y)}{\partial d_{11}} +
             (d_{11}+d_{22})
             \frac{\partial \gamma_2(y)}{\partial d_{11}} +
             \gamma_2(y)+y\gamma_1(y)
           \right )
   \label{der-m2-d11}\\
   & & \cr
\frac{\partial m_2(y)}{\partial d_{20}} &=& 
           2\left (
             (d_{21}+y(d_{11}+2d_{22}))
             \frac{\partial \gamma_1(y)}{\partial d_{20}} +
             (d_{11}+d_{22})
             \frac{\partial \gamma_2(y)}{\partial d_{20}} +
             1
           \right )
   \label{der-m2-d20}\\
   & & \cr
\frac{\partial m_2(y)}{\partial d_{21}} &=& 
           2\left (
             (d_{21}+y(d_{11}+2d_{22}))
             \frac{\partial \gamma_1(y)}{\partial d_{21}} +
             (d_{11}+d_{22})
             \frac{\partial \gamma_2(y)}{\partial d_{21}} +
             \gamma_1(y)+y
           \right )
   \label{der-m2-d21}\\
   & & \cr
\frac{\partial m_2(y)}{\partial d_{22}} &=& 
           2\left (
             (d_{21}+y(d_{11}+2d_{22}))
             \frac{\partial \gamma_1(y)}{\partial d_{22}} +
             (d_{11}+d_{22})
             \frac{\partial \gamma_2(y)}{\partial d_{22}} +
             2y\gamma_1(y)+\gamma_2(y)+y^2
           \right )
   \label{der-m2-d22}
\end{eqnarray}
\label{der-gamma-mm}
\end{subequations}
where
\begin{subequations}
\begin{eqnarray}
h_1(y) &=& \int_{-\infty}^{+\infty} (x^{\prime}-y)^3 \bar{f}_{\sigma}(x^{\prime}\vert y) dx^{\prime} 
    \label{h1}\\
    & & \cr
h_2(y) &=& \int_{-\infty}^{+\infty} (x^{\prime}-y)^4 \bar{f}_{\sigma}(x^{\prime}\vert y) dx^{\prime} .
    \label{h2}
\end{eqnarray}
\label{hmore}
\end{subequations}

\end{widetext}


\end{document}